\documentclass[aps,prb,reprint,groupedaddress,showpacs,dvipdfmx]{revtex4-1}
\usepackage{graphicx}
\usepackage{dcolumn}
\usepackage{bm}
\usepackage{amsmath}
\usepackage{amssymb}
\usepackage{txfonts}
\usepackage{multirow}
\usepackage{url}
\usepackage{here}
\usepackage{braket}
\usepackage[usenames,dvipsnames]{color}
\definecolor{Gray}{gray}{0.0}
\definecolor{lightGray}{gray}{0.35}
\usepackage{xr}
\newcommand{\tvb}{\textsc{TurboRVB}} 
\newcommand{\enumi}{{$\rm(\hspace{.18em}i\hspace{.18em})$ }}
\newcommand{\enumii}{{$\rm(\hspace{.08em}ii\hspace{.08em})$ }}
\newcommand{\enumiii}{{$\rm(i\hspace{-.08em}i\hspace{-.08em}i)$ }}

\newcommand{\kosuke}[1]{{\rm{#1}}}

\begin{document}
\title{Atomic forces by quantum Monte Carlo: Application to phonon dispersion calculations}
\author{Kousuke Nakano$^{1,2}$}
\email{kousuke\_1123@icloud.com}

\author{Tommaso Morresi$^{3}$}

\author{Michele Casula$^{3}$}

\author{Ryo Maezono$^{2}$} 

\author{Sandro Sorella$^{1}$}
\email{sorella@sissa.it}
  
\affiliation{$^{1}$ International School for Advanced Studies (SISSA), Via Bonomea 265, 34136, Trieste, Italy}

\affiliation{$^{2}$ Japan Advanced Institute of Science and Technology (JAIST), Asahidai 1-1, Nomi, Ishikawa 923-1292, Japan}

\affiliation{$^{3}$ Institut de Min{\'e}ralogie, de Physique des Mat{\'e}riaux et de Cosmochimie (IMPMC), Sorbonne Universit{\'e}, CNRS UMR 7590, IRD UMR 206, MNHN, 4 Place Jussieu, 75252 Paris, France}

\date{\today}

\begin{abstract}
We report a successful application of the {\it ab initio} quantum Monte Carlo (QMC) framework to a phonon dispersion calculation. A full phonon dispersion of diamond is successfully calculated at the variational Monte Carlo (VMC) level, based on the frozen-phonon technique. 
The VMC-phonon dispersion is in good agreement with the experimental results, 
giving renormalized harmonic optical frequencies very close to the experimental values,
{\kosuke {and improving upon previous density functional theory estimates.}}
Key to success for the QMC approach is the statistical error reduction in the atomic force evaluation. We show that this can be achieved by using well conditioned atomic basis sets and by explicitly removing the basis-set redundancy, which reduces the statistical error of forces by up to two orders of magnitude
{\kosuke{by combining it with the so-called space-warp transformation algorithm.}}
This leads to affordable and accurate QMC-phonons calculations, 
which are up to $10^{4}$ times more efficient than {\kosuke{a bare force treatment,}}
and paves the way to new applications, particularly in correlated materials, where phonons have been poorly reproduced so far.
\end{abstract}
\maketitle

\makeatletter
\def\Hline{
\noalign{\ifnum0=`}\fi\hrule \@height 1pt \futurelet
\reserved@a\@xhline}
\makeatother


The accurate description of 
phonons
in a solid is one of the central research topics 
in the field of condensed matter physics and 
materials science for discussing phase stability (i.e., Gibbs-free energy), 
electron-phonon interaction, and structural phase transitions of materials~{\cite{2004MAR, 2011DAV}}.
{\it Ab initio} phonon calculations based on the Density Functional Theory (DFT)~\cite{2001BAR, 2015TOG} have been successful for many compounds,
but they often fail in strongly-correlated materials.
For example, 
DFT calculations severely underestimate the highest frequency of the optical phonons of graphene at the $K$ point~{\cite{2005MOU, 2008LAZ}}, because the  electron-electron correlation  is not  taken into account with sufficient accuracy. 
Another example is the elemental cerium, whose phonon dispersions measured by neutron scattering  strongly mismatch with the calculated DFT-PBE ones~\cite{2011KRI}.
{\kosuke{Interestingly, such failure was also seen in a high-$T_c$ cuprate superconductor~{\cite{2008REZ}}}
}
Some effort has been made to include correlation in phonon calculations in the DFT+U framework~\cite{2011FLO} and also within the DMFT framework~\cite{2012LEO,2014LEO}. In both cases, this requires modeling correlations by an {\it empirical parameter}, though physically motivated (i.e., the Hubbard $U$). Indeed, a {\it genuine ab initio} framework applicable to strongly correlated materials without any empirical parameters remains, so far, 
a very important theoretical challenge.

\vspace{1mm}
The {\it ab initio} quantum Monte Carlo (QMC) framework, 
including variational quantum Monte Carlo (VMC) and the diffusion quantum Monte Carlo (DMC) schemes, is among the state-of-art numerical methods to obtain highly accurate many-body wave functions{~\cite{2001FOU}}, and cope with the electron correlation more rigorously than DFT.
It has been successfully applied to challenging materials that DFT cannot tackle, such as cuprates~\cite{2011MAR}, iron-arsenides~\cite{2013CAS}, and graphene~\cite{2011MAR, 2018SOR}.
So far, unfortunately,  almost all QMC applications are mainly based on energy and its first derivative (i.e., atomic force) calculations.
Indeed, it is at present  an open  problem how to evaluate,  
 with an affordable computational effort, second and higher-order derivatives, which are essential for computing various physical properties. 

\vspace{1mm}
There are three routes to compute the second derivatives (i.e., $\frac{\partial^{2}E}{\partial {R_{\alpha}}\partial {R_{\beta}}}$), which are needed for 
evaluating {\it harmonic} phonon properties,  by {\it ab initio} calculations, i.e., 
\enumi
potential energy surface (PES) fitting, 
\enumii
finite difference expression based on atomic force evaluations
, and
\enumiii
direct evaluation of second derivatives.
All of the above attempts have been successful for isolated molecular systems~\cite{2012ZEN,2014LUO,2019LIU,2019NAK}.
On the other hand, for solids, only strategy \enumi has been successful so far within the QMC framework. For example, Maezono {\it et al.} calculated Raman frequencies of diamond (phonons at $q=\Gamma$ point)~{\cite{2007MAE}}. 
However, QMC-phonon calculations of solids have been limited to a single high-symmetry $q$-point. To the  best of our knowledge, 
the full ($q$-resolved) phonon dispersion has been unaccessible so far at both VMC and DMC levels.

\vspace{1mm}
In this paper, we report a successful phonon dispersion calculation of diamond at the VMC level by adopting strategy \enumii, the so-called frozen phonon technique~\cite{2015TOG}.
The key to success is to reduce the statistical error of atomic forces.
We found that removing the nearly linear dependency  of the basis set used for 
the trial wave function parametrization~\cite{2010AZA} dramatically 
lowers the statistical error of forces.
Its decrease reaches the order of $\times$ ${\sim 10^{-2}}$, which corresponds to a speed-up of ${\sim 10^{4}}$ times in a VMC computation. 
This drastic reduction enables us to construct a dynamical matrix within an affordable computational time, and to eventually apply VMC-phonon calculations to 
new interesting materials from first principles.

\begin{center}
\begin{table*}[htbp]
\caption{\label{Comp_ph_DFT_QMC} The highest harmonic phonon frequencies (THz)  of diamond at high-symmetry $q$ points. All phonons are calculated using the experimental lattice parameter. In this table, (P) denotes the interpolated frequencies obtained by Phonopy,
(F) denotes the phonon frequency at $\Gamma$ obtained by fitting forces of distorted structures along the Raman mode with a linear function at the VMC level,
and (E) denotes the phonon frequency at $\Gamma$ obtained by fitting energies of undistorted and distorted structures along the Raman mode with a quadratic function at the VMC and LRDMC levels. The last column indicates the harmonic frequencies estimated by the raw experimental values:
 39.938~THz\cite{2000LIU}, 35.299~THz\cite{1998SCH}, and 37.962~THz\cite{2002KUL} at the $\Gamma$, $X$, and $L$ points, respectively.} 
\vspace{2mm}
\begin{tabular}{c|c|c|cc|cccc|c}
\Hline
\multirow{2}{*}{$q$} & \multicolumn{2}{c|}{DFT} & \multicolumn{2}{c|}{Previous work} & \multicolumn{4}{c|}{This work{\footnotemark[4]}} & Experiment \\
\cline{2-10}
 & LDA-PZ  &  GGA-PBE   & VMC{\footnotemark[3]}      & DMC{\footnotemark[3]}       & VMC(P)      & VMC(F) & VMC(E) & LRDMC(E)     & Harmonic (Estimated) \\ 
\hline
$\Gamma$ & 38.55{\footnotemark[1]} & 38.82{\footnotemark[2]} & 41.64(9) & 41.22(12) & 40.65(38) & 40.49(4) & 40.68(29) & 41.52(22) & 40.460{\footnotemark[5]}, 40.349{\footnotemark[6]} \\ \hline
$X$      & 35.64 & 35.87 & -        & -         & 36.48(40) & -    & -    & -     &  35.476{\footnotemark[6]}  \\ \hline
$L$      & 37.31 & 37.47 & -        & -         & 38.01(31) & -    & -    & -     & 38.242{\footnotemark[6]} \\ \hline
\Hline
\end{tabular}
\footnotetext[1]{38.40 THz in the previous DFT study employing the LDA functional. See Ref.~{\onlinecite{2007MAE}}.}
\footnotetext[2]{38.73 THz in the previous DFT study employing the GGA-PBE functional. See Ref.~{\onlinecite{2007MAE}}.}
\footnotetext[3]{These values are taken from Ref.~{\onlinecite{2007MAE}}.}
\footnotetext[4]{These values include the one-body finite size corrections, i.e., $-$ 0.16 THz, $-$ 0.18 THz, and $-$ 0.23 THz for $q$ = $\Gamma$, $X$, and $L$, respectively.}
\footnotetext[5]{The anharmonic renormalization, $-$ 17.4 cm$^{-1}$ = $-$ 0.522 THz was employed. See Ref.~{\onlinecite{1984VAN}}.}
\footnotetext[6]{The anharmonic renormalizations are $-$ 0.411 THz, $-$ 0.177 THz, and $-$ 0.280 THz for $\Gamma$, $X$, and $L$, respectively, which were estimated by molecular dynamics simulations performed in this work.}
%
\end{table*}
\end{center}



\begin{figure}[htbp]
  \centering
  \includegraphics[width=8.6cm]{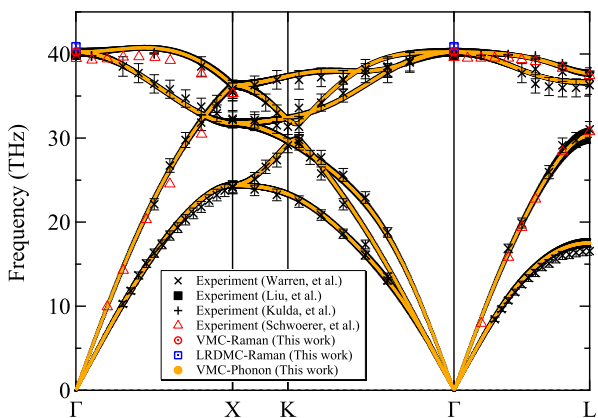}
  \caption{The phonon dispersion and Raman frequencies of diamond calculated using a 2 $\times$ 2 $\times$ 2 conventional supercell at the VMC level, where the anharmonic and one-body finite size corrections are included.
The experimental lattice parameter (i.e., 6.741 bohr~{\cite{1960PAR}}) was employed. The error bars were estimated by the jackknife method. {\it Observed} experimental results are also plotted for comparison.
Experimental data are taken from Refs.~{\onlinecite{1965WAR, *1967WAR, 2000LIU, 2002KUL, 1998SCH}}. 
The data in Ref.~{\onlinecite{1998SCH}} were digitalized using WebPlotDigitizer~\cite{2020ROH}.
  }
  \label{pd_disp_diamond}
\end{figure}


\vspace{1mm}
All Variational Monte Carlo (VMC) and lattice regularized diffusion Monte Carlo (LRDMC)~{\cite{2005CAS, 2020NAK1}} calculations in this study were performed by 
the \tvb\ ~{\cite{2003CAS, 2020NAK2}} SISSA quantum Monte Carlo package.
We employed the Jastrow Slater-determinant(JSD) ansatz, 
i.e,
$
\Psi  =  \Psi _\text{SD} \times \exp J,
$
where $\Psi _\text{SD}$ and $J$ are 
the Slater determinant and Jastrow terms, respectively.
The Slater determinant part is expressed in terms of molecular orbitals $\phi_k(\mathbf{r}) = \sum_{i=1}^{L}{c_{i,k}{\psi_{i}}(\mathbf{r})}$ expanded in a 
periodized Gaussian
basis set $\{\psi_{i} \}$.
The valence triple-zeta (VTZ) basis set accompanying an energy-consistent pseudo potential developed by Burkatzki {\it et al.}~{\cite{2007BUR}} was employed for the primitive Gaussian atomic orbitals (Table~\ref{basis_set}{\cite{foot}}).
The coefficients of atomic orbitals (i.e., $c_{i,k}$) were obtained by a DFT calculation {\kosuke{with the LDA-PZ exchange-correlation functional}~\cite{1981PER}} and were left unchanged during the VMC optimization.
The Jastrow 
factor
was composed of inhomogeneous one-, two- and three-body  
contributions
($J = {J_1}^{inh}+{J_2}+{J_{3}}$).~\cite{foot}
The variational parameters in the Jastrow terms 
were optimized by the stochastic reconfiguration~\cite{2007SOR} and/or the modified linear method~\cite{2007UMR, 2020NAK2} implemented in \tvb.
LRDMC calculations were performed by the single-grid scheme~\cite{2005CAS} with a lattice space, $a$ = 0.2 bohr.

\vspace{1mm}
In this paper, we focus on diamond (Space group: $Fd\bar{3}m$) as a proof of concept for the first example of a VMC-based phonon dispersion calculation. 
2 $\times$ 2 $\times$ 2 conventional supercells (256 electrons / 64 carbon atoms in a simulation cell) were used for most calculations, while 3 $\times$ 3 $\times$ 3 conventional supercells (864 electrons / 216 carbon atoms in a simulation cell) were also used for several calculations to investigate the finite size errors.
$L$-twist (i.e., $k$ = $\pi$, $\pi$, $\pi$) was employed for alleviating the so-called one-body finite-size effects~\cite{2007MAE, 2010HEN, 2011SOR}.
Dynamical matrices and the corresponding phonon dispersions were calculated based on the frozen-phonon technique implemented in the Phonopy module~{\cite{2015TOG}},
where a 0.15 bohr displacement of carbon atoms was large  enough to work with reasonable signal/noise ratio in QMC forces. 
This displacement underestimates harmonic frequencies only by $\sim$ 0.1~THz, as shown in Fig.~\ref{fdm-dist-freq-conv}~\cite{foot}.
Error bars in a phonon dispersion were estimated by the jackknife method~{\cite{2017BEC}}. 
Phonon calculations based on DFT were performed using the \textsc{Quantum Espresso} package~\cite{2009GIA} with LDA-PZ~\cite{1981PER} and GGA-PBE~\cite{1996PER} exchange-correlation functionals at the experimental lattice parameter.
~\cite{foot}
The phonon dispersion of diamond has already been 
studied using DFT calculations by many groups so far, at the theoretical ~\cite{1993PAV, 1995KRE} and experimental~\cite{2007MAE} lattice parameters, that makes this system a very good testbed for any new methodological implementation of phonons calculations.
As shown later, our DFT calculations are consistent with the previous study.

\begin{center}
\begin{table*}[htbp]
\caption{\label{Table2} Vinet EOS parameters. Zero point energy (ZPE) and temperature effects (TE) are not included in these theoretical data.}

\begin{tabular}{c|c|c|cc|cc|c|c}
\Hline
\multirow{2}{*}{Parameter} & \multicolumn{2}{c|}{DFT} & \multicolumn{4}{c|}{QMC} & \multicolumn{2}{c}{Experiment} \\
\cline{2-9}
 & LDA-PZ &  GGA-PBE  & VMC{\footnotemark[1]}      & DMC{\footnotemark[1]}       & VMC       & LRDMC     & w/o ZPE, w/o TE & Observed \\ 
\hline
Lattice (Bohr)       & 6.683 & 6.748 & 6.691(4) & 6.734(4)  & 6.693(1) & 6.702(1) & 6.7193(5){\footnotemark[2]}  & 6.7410(5){\footnotemark[3]}           \\
$V_{0}$ (bohr$^3$)   & 37.32 & 38.41 & 37.45(6) & 38.17(6)  & 37.47(2) & 37.63(2) & 37.920(9){\footnotemark[2]} & 38.290(9){\footnotemark[3]}          \\
$B_0$   (GPa)        & 465   & 433   & 483(4)   & 448(3)    & 476(6)   & 463(5) & 457(1){\footnotemark[2]}, 453(5){\footnotemark[2]} & 446(1){\footnotemark[3]}, 442(5){\footnotemark[4]}           \\
$B'_0$               & 3.65  & 3.70  & 3.8(1)   & 3.7(1)    & 4.0(6)   & 4.9(6)  & 3.0(1){\footnotemark[2]}, 4.0(7){\footnotemark[2]} & 3.0(1){\footnotemark[3]}, 4.0(7){\footnotemark[4]}\\
\Hline
\end{tabular}

\footnotetext[1]{These values are taken from Ref.~{\onlinecite{2007MAE}}. Here, ZPE and TE are subtracted.}
\footnotetext[2]{ZPE and TE are corrected, $-$ 0.37 Bohr$^3$, $+$ 11 GPa, and $-$ 0.03 for $V_{0}$, $B_0$, and $B'_0$, respectively. See Ref.~{\onlinecite{2007MAE}}.}
\footnotetext[3]{These values are taken from Ref.~{\onlinecite{2003OCC}}.}
\footnotetext[4]{These values are taken from Refs.~{\onlinecite{1972MCS,2003OCC}}.}
\end{table*}
\end{center}

\begin{figure}[htbp]
  \centering
  \includegraphics[width=8.6cm]{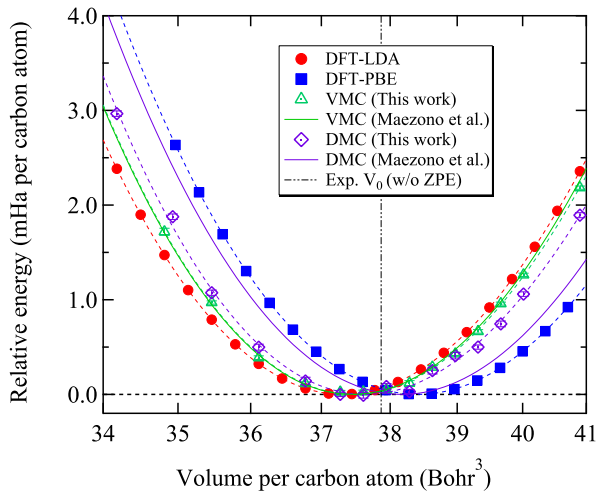}
  \caption{Equation of states and the curves fitted by the Vinet exponential function~\cite{1987VIN}. VMC and DMC results~(Maezono et al.) are taken from Ref.~{\onlinecite{2007MAE}}.
  }
  \label{eos}
\end{figure}

\begin{figure*}[htbp]
  \centering
  \includegraphics[width=17.6cm]{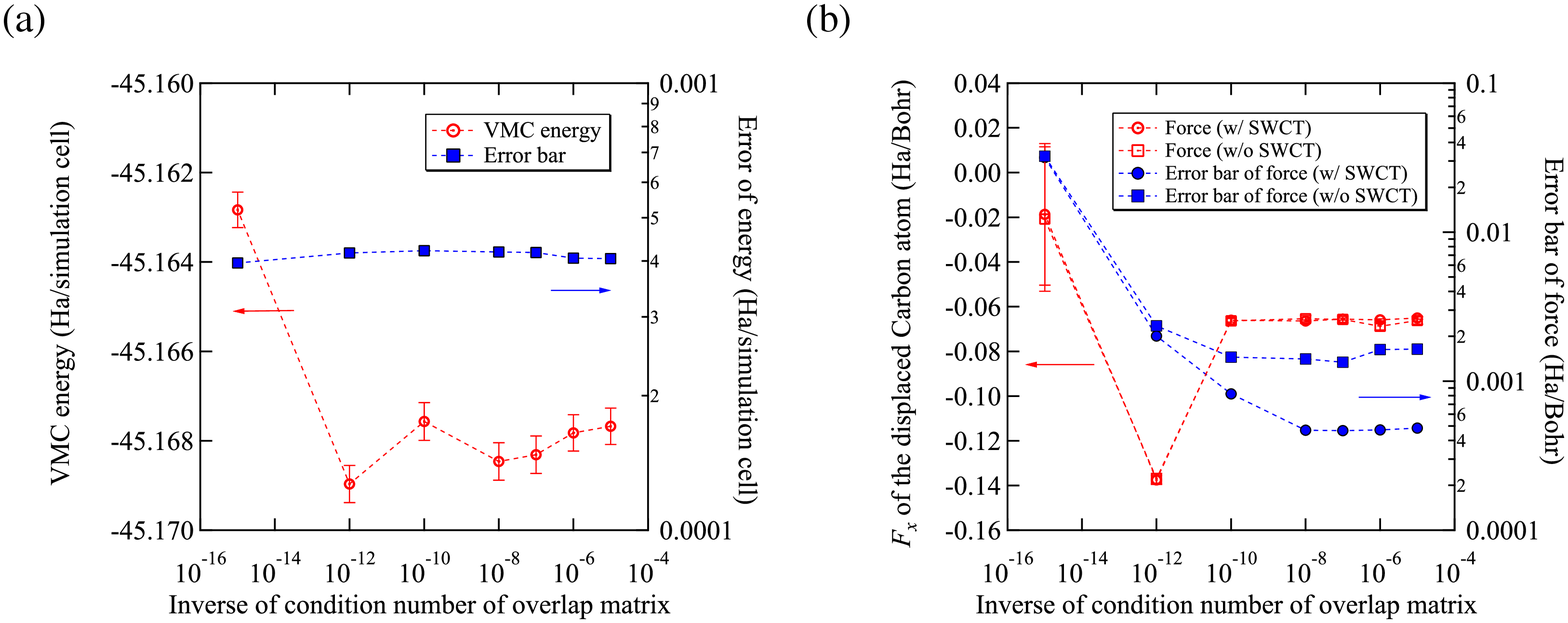}
  \caption{(a) VMC energies and their statistical error bars and (b) VMC forces and their statistical error bars vs. inverse of the condition number of the overlap matrix. The numerical data is shown in Table~{\ref{force_errors}}{~\cite{foot}}. The forces were calculated with and without the space warp coordinate transformations (SWCT)~\cite{2010SOR}. The Hellmann-Feynman and Pulay contributions to the forces are shown in Fig.~\ref{err_force}~\cite{foot}. The same VTZ atomic basis set and the pseudo potential~{\cite{2007BUR}} as in the phonon and EOS calculations were adopted, while 1 $\times$ 1 $\times$ 1 conventional cell (8 carbon atoms in a simulation cell) with $k$ = $\Gamma$ twist was employed. Only one carbon atom was displaced in $x$ direction by 0.15 bohr (Table~{\ref{displaced_diamond}}){~\cite{foot}}.}
  \label{epsover}

\end{figure*}


\vspace{1mm}
Fig.~{\ref{pd_disp_diamond}} shows the phonon dispersion 
obtained by our VMC calculations using the conventional 
2 $\times$ 2 $\times$ 2 diamond supercell
{\footnote{
2 $\times$ 2 $\times$ 2 conventional supercell containing 64 atoms in the simulation cell is large enough for obtaining a phonon dispersion almost consistent with experimental ones, as shown in Fig.~\ref{one-body-dft}~\cite{foot}.
This was confirmed by comparing the phonon dispersion obtained by the finite displacement method and that obtained by the linear-response method with very dense $k$ and $q$ grids at the DFT level.
}}.
{\it Observed} experimental frequencies~\cite{1967WAR, 2000LIU, 2002KUL} are also plotted for comparison.
A phonon dispersion obtained by the finite displacement method does not include anharmonic effects, which can decrease harmonic frequencies by up to $\sim$ 5-10\% for the lightest elements~\cite{1995HAR}. 
Therefore, for comparison with the experimental results, anharmonic corrections were added to the VMC phonon dispersion in this study. The phonon dispersion before the correction is shown in Fig.~{\ref{pd_disp_diamond_diff}}~\cite{foot}
The anharmonic renormalizations were estimated in this study using path integral molecular dynamics simulations\cite{2017MOU,2020MOR} {\kosuke{at the DFT level with the PBE exchange-correlation function (Fig.~{\ref{pimd_spectrum}}~\cite{foot})}}, giving $-$0.411 THz, $-$0.177 THz, and $-$0.280 THz for $\Gamma$, $X$, and $L$, respectively (see the SI for details). The value at $\Gamma$ is consistent with a reported estimate of
$-$17.4 cm$^{-1}$ = $-$0.522 THz ~{\cite{1984VAN}}.
Notice that other possible sources of error were also considered.
The phonon dispersion in Fig.~{\ref{pd_disp_diamond}} also includes the one-body finite size corrections that were estimated by DFT calculations (see Fig.~{\ref{one-body-dft}}~\cite{foot})
The two-body finite-size error is negligible because the average density and the volume do not change in phonon calculations.
Table~\ref{Comp_ph_DFT_QMC} shows a detailed comparison of the highest harmonic phonon frequencies at three $q$ points, i.e., $\Gamma$ = (0,0,0), $X$ = (2$\pi$,0,0), and $L$ = ($\pi$,$\pi$,$\pi$). 
Raman frequencies at $\Gamma$ obtained by a direct fit of
VMC energies, VMC forces, and LRDMC energies of the structures displaced along the corresponding eigenmode are also plotted in Fig.{~\ref{pd_disp_diamond}}.
{\footnote{
In the Raman mode, two nearest-neighbor carbon atoms are displaced in opposite directions by a distance $u$ from their high-symmetry positions~\cite{2007MAE}.
The distortion in the [1 0 0] direction was employed in this study~\cite{1984VAN}.
We calculated VMC forces for two distorted structures ($u = 0.03$ Bohr and $u = 0.05$ Bohr) and obtained the Raman frequency by fitting the forces with a linear function. We also calculated VMC and LRDMC energies for the undistorted structure and two distorted structures ($u = 0.03$ Bohr and $u = 0.05$ Bohr), then obtained Raman frequencies by fitting the energies with a quadratic function.
}}
In Table~\ref{Comp_ph_DFT_QMC} and hereafter, (P) denotes the interpolated frequencies obtained by Phonopy,
(F) denotes the phonon frequency at $\Gamma$ obtained by 
force fitting,
and (E) denotes the phonon frequency at $\Gamma$ obtained by 
energy fitting.
The corresponding Raman frequencies are consistent within the error bars,
indicating that the Slater determinant obtained by DFT is almost optimal also in the 
presence of the Jastrow factor, 
\kosuke{thus explaining why this consistency is satisfied quite accurately~{\cite{2014MOR}}.
In other words, if the wavefunction is at its minimum, the consistency
is a consequence.}
\kosuke{If it is not at its minimum, the consistency may also be satisfied by chance or good behavior of the used basis set.}

\vspace{1mm}
All VMC(P), VMC(F), and VMC(D) calculations give the harmonic phonon frequency at $\Gamma$ very close to the experimental value, considering the renormalization of the anharmonic effect, i.e., the discrepancy is just $\sim$ 0.3~THz.
On the other hand, both DFT calculations with LDA-PZ and GGA-PBE exchange-correlation functionals underestimate the highest frequency at $\Gamma$ by $\sim$ 1.8~THz and $\sim$ 1.5~THz, respectively. Table~\ref{Comp_ph_DFT_QMC} shows that our VMC phonon dispersion calculation also gives accurate harmonic frequencies at other $q$ points, i.e., at $X$ and $L$. Compared with the previous VMC study~\cite{2007MAE}, our VMC frequency at $\Gamma$ point is closer to the experimental value, as reported in Table~\ref{Comp_ph_DFT_QMC}.
{\footnote
{Since smaller displacements were employed in this study (i.e., 0.03 and 0.05 Bohr) than the previous one, the error bars in phonon frequencies at $\Gamma$ point are a bit larger.}
}
The improvement at the VMC level certainly derives from the more flexible Jastrow factor 
employed in this study, while the one used in the previous VMC study was much simpler~\cite{2004DRU, 2007MAE}.

\vspace{1mm}
It is intriguing that our LRDMC calculation gives a 1.2(2)~THz higher Raman frequency than the experimental one, as shown in Table~\ref{Comp_ph_DFT_QMC}.
The previous DMC study also overestimated the Raman frequency by 0.9(1)~THz~\cite{2007MAE}.
To discuss the origin of the discrepancy, we investigated the effect of the lattice-space error in our LRDMC calculations. 
An extrapolation ($a \to 0$) with four lattice spaces 
(i.e., 0.20, 0.30, 0.40, and 0.50 bohr) 
yields 41.89(44) THz,
suggesting that the lattice-space error 
is not the origin of the discrepancy.
We also suspected that the experimental lattice parameter employed in the phonon calculation could be significantly different from the theoretical one, 
but this is also not the origin, as shown later.
Therefore, the discrepancy should arise from the fixed-node error
and this should be alleviated by a nodal surface optimization~\cite{2020NAK2}, which however is prohibitive in the 2 $\times$ 2 $\times$ 2 conventional supercells {\kosuke{due to the large number of variational parameters of the distorted structures}}.
{\kosuke{A possible future work to study the nodal surface effect is the use of various exchange-correlation functionals suitable for solids such as HSE06~\cite{2006KRU} and SCAN~\cite{2015SUN}}


\vspace{1mm}
Since the equilibrium lattice parameter also affects phonon frequencies, we 
investigated the equation of states (EOSs) of diamond.
Figure~\ref{eos} shows plots of internal energies vs. volumes 
fitted by the Vinet EOS~\cite{foot}
Previous VMC and DMC results~\cite{2007MAE}, and the experimentally observed equilibrium lattice parameter are also plotted in Fig.~\ref{eos} and summarized in Table~\ref{Table2}. In Fig.~\ref{eos}, the zero point energy (ZPE) contribution\cite{2007MAE,2012HAO} is subtracted, to make the comparison possible with internal energies computed
at $T=0$ and on static lattice.
%
%
Table~\ref{Table2} shows that our VMC calculation reproduces the previous VMC study, while our LRDMC calculation gives a slightly smaller lattice parameter (6.702(1) bohr) than the previous DMC study (6.734(4) bohr). This discrepancy 
likely
derives from the different nodal surfaces used in the two studies, namely the one originating from the DFT-PBE orbitals in Ref.~\onlinecite{2007MAE} and the one coming from the LDA orbitals in our work.
Notice that both one-body and two-body finite size errors are negligible for the EOS calculations as shown in Fig.~{\ref{finite-size-corr}}~\cite{foot}
%
Table~\ref{Table2} indicates that the equilibrium lattice parameter of diamond is $\sim$ 0.03 bohr ($\sim$ 0.02 bohr) smaller than the experimental one at the VMC (LRDMC) level. Our DFT calculations (Fig.~\ref{Raman_freq_lattice_dep}~\cite{foot})
suggest that the Raman frequency is implicitly proportional to the lattice parameter, with a
gradient of $-$0.18~THz/0.01~bohr.
Therefore, if the equilibrium lattice parameter (with ZPE)
were employed instead of the experimental one,
VMC (LRDMC) calculations would give
$\sim$ 0.5 THz ($\sim$ 0.4 THz) higher Raman frequencies,  
while still staying close to the experimental values.

\vspace{1mm}
Reducing the statistical errors of atomic forces is key to a successful phonon calculation. We found that alleviating the linear dependency of a localized atomic basis set drastically decreases the statistical error.
In general, a basis set optimized for molecular systems is not suitable for solid state calculations, 
{\kosuke{(i.e., strongly linear dependent) due to the presence of orbitals having small exponents (c.f., typically $<$ 0.1)~\cite{2013PEI}.}}
The quality of the basis set is systematically improved by a general and efficient scheme implemented in \tvb~{\cite{2010AZA}}.
Indeed, the linear dependency of a localized atomic basis set ($\psi_{\mu}(\vec{r})$) is characterized by the condition number, $\kappa ({\bf S})$, where $S_{\mu,\nu} = \braket{\psi_{\mu}|\psi_{\nu}}$ is the overlap matrix~{\footnote{$\kappa ({\bf S})$ = the maximum eigenvalue / the minimum eigenvalue of the overlap matrix ($S_{\mu,\nu} = \braket{\psi_{\mu}|\psi_{\nu}}$)}}.
In \tvb, a redundant basis set is converted into a well-conditioned one, by disregarding small eigenvalues and the corresponding eigenvectors of the overlap matrix ${\bf S}$~{~\cite{2010AZA, 2020NAK2}}.
{\kosuke{We note that a well-conditioned basis set can also be constructed by simply removing orbitals having small exponents, while the method employed in this work is more general and systematic.}}
Figure~{\ref{epsover}} shows the plots of VMC energies, VMC forces, and their statistical errors vs. the inverse of the condition number ($\kappa ({\bf S})^{-1}$).
Figure~{\ref{epsover}} (a) indicates that the statistical error of the energy is independent of the condition number, at variance with the statistical error of the force, which instead strongly depends on it (Fig.~{\ref{epsover}} (b)).
The error bar in the forces amounts to {\kosuke{$\sim 3.4 \times 10^{-2}$ (Ha/bohr)} when the atomic basis set is strongly linear dependent (i.e., $\kappa ({\bf S})^{-1} = 10^{-15}$), a condition that also introduces some bias in the forces
because, as we have verified, they are no longer consistent with the finite difference energy derivatives for $\kappa ({\bf S})^{-1} < 10^{-12}$ (see Fig.~{\ref{epsover}}b)
On the other hand, the statistical error becomes much smaller {\kosuke{$\sim 1.7 \times 10^{-3}$ (Ha/bohr)}} by removing the linear dependency (i.e., $\kappa ({\bf S})^{-1} = 10^{-7}$).
{\kosuke{The space warp coordinate transformation (SWCT)~\cite{2010SOR} is able to further reduce the statistical error. Indeed, Fig.~{\ref{epsover}} (b) shows that the error bar of the force becomes $\sim 4.1 \times 10^{-4}$ (Ha/bohr) by using the SWCT algorithm combined with a well-conditioned basis set, corresponding to $\sim 10^{4}$ times more efficient computation than a bare force treatment.}}

\vspace{1mm}
We analyze now in detail the reason of this behavior.
\tvb\ evaluates atomic forces in a differential form (i.e, by the algorithmic differentiation)~\cite{2008ATT,2010SOR}:
\begin{equation*}
{F_\alpha } = {\braket {\frac{{dE}}{{d{R_\alpha }}}}} _ {\left| {\Psi}^{\bf{R}} \right|^2} 
\simeq 
{\braket
{\frac
{E\left({\bf{R}} + \Delta {\bf{R}_{\alpha}} \right) - E\left( {\bf{R}} \right)}
{{\Delta {R_\alpha }}}
}
}_{\left| {\Psi}^{\bf{R}} \right|^2},
\end{equation*}
where ${\bf{R}} = (R_1, \ldots, R_{\alpha}, \ldots)$ and $\Delta {\bf{R}_{\alpha}} = (0, \ldots, \Delta R_{\alpha}, \ldots)$.
This equation suggests
that the statistical error on forces depends
on how much the wavefunction changes
after an atom is displaced. 
In other words, to minimize the stochastic error,
the overlap $\braket{\Psi ^{{\bf{R}} + \Delta {\bf{R_{\alpha}}}} | \Psi ^{\bf{R}}}$ should be close to unity.
To investigate the effect of the linear dependency on the overlap, we calculated $\braket{\Psi ^{{\bf{R}} + \Delta {\bf{R}}_{\alpha}} | \Psi ^{\bf{R}}}$ with linear-dependent and linear-independent basis sets, using correlated sampling techniques~\cite{2020NAK2}, where only one carbon atom in the 1 $\times$ 1 $\times$ 1 conventional simulation cell was displaced in the $x$ direction by $\Delta R_\alpha = 0.005$ bohr.
We obtained 0.9999 and 0.9726 for $\kappa ({\bf S})^{-1} = 10^{-7}$ and $\kappa ({\bf S})^{-1} = 10^{-15}$, respectively. This clearly indicates that the linear dependency of the basis set deteriorates the overlap $\braket{\Psi ^{{\bf{R}} + \Delta {\bf{R}}_{\alpha}} | \Psi ^{\bf{R}}}$, 
thus increasing the statistical error of forces.

\vspace{1mm}
The deterioration is explained as follows:
Here, a simple Slater wavefunction without Jastrow factor is considered for the sake of clarity. In this case, the overlap $\braket {\Psi ^{{{\bf{R}}}} | {\Psi ^{{{\bf{R}} }}} }$ reads 
$
\braket {\Psi ^{{{\bf{R}} }} | {\Psi ^{{{\bf{R}} }}} } ={\rm det}  \langle \phi_i^{\bf{  R}} | \phi_j^{\bf{  R} } \rangle,
$
where $\phi_i^{\bf{R}}$ is the $i$-th molecular orbital depending on 
nuclear positions {\bf{R}}, defining the above $N \times N$ determinant matrix. The molecular orbital is expanded over localized atomic orbitals, i.e., 
$\phi_i^{\bf{R}} = \sum_{a,l}{ c_{i,\{a,l\}}} \psi_{\{a,l\}}^{R_a}$,
where $\psi_{\{a,l\}}^{R_a}$ are (periodized) atomic orbitals explicitly dependent on a 
nuclear position $R_{a}$ $\{ a = 1, \ldots, \alpha, \ldots \}$, while $c_{i,\{a,l\}}$ are nuclear position independent.
We can readily derive $\braket {\Psi ^{{{\bf{R}}}} | {\Psi ^{{{\bf{R}} }}} } = 1$ when the molecular orbitals are orthonormalized (i.e., $\braket {\phi _{i}^{{{\bf{R}} }} | \phi _{j}^{{{\bf{R}}}}} = \delta_{i,j}$).
What about $\braket {\Psi ^{{{\bf{R}} + \Delta {\bf{R_{\alpha}}}}} | {\Psi ^{{{\bf{R}}}}}}$?
We would like to show how the 
perturbation 
$\frac{d}{d_{R_{\alpha}}} \phi _{i} ^{\bf{R}} \equiv \sum_{a=\alpha, l}{ c_{i,\{a,l\}}} \frac{d}{d_{R_{\alpha}}} \psi_{\{a,l\}}^{R_a}$ affects the overlap.
In DFT calculations, it turns out  that $\left| c_{i,\{a,l\}} \right| \gg 1$ 
when the basis is redundant (e.g., $\kappa ({\bf S})^{-1} = 10^{-15}$), while $\left| c_{i,\{a,l\}} \right| \ll 1$ when the basis set is well-conditioned (e.g., $\kappa ({\bf S})^{-1} = 10^{-7}$).
When $\left| c_{i,\{a,l\}} \right| \ll 1$, the perturbation effect is rather small, 
and
the orthonormalization condition almost holds, while $\left| c_{i,\{a,l\}} \right| \gg 1$ makes the perturbation effect significant, and by consequence
the orthonormalization condition is certainly deteriorated. 
This is why the linear dependency of an atomic basis set deteriorates the overlap and, thus, induces a large error bar in forces. Thus, a complete all-electron and pseudopotential basis set database suitable for QMC {\kosuke{{\it solid state}}} calculations will be quite useful for the application of QMC-forces to the calculations of phonons in realistic materials.


%
\vspace{1mm}
In summary, we report a 
VMC determination of the momentum-resolved phonon dispersion of diamond.
Our approach combines the {\it ab initio} quantum Monte Carlo
framework with the so-called frozen phonon technique. It gives results
in very good agreement with experiments and provides renormalized
harmonic optical frequencies consistent with the experimental
findings. We estimated the purely harmonic contribution to the phonon
spectrum, by evaluating $q$-dependent anharmonic corrections by means
of a path integral molecular dynamics driven by
PBE forces\cite{2020MOR}. After including these corrections, the VMC
phonon spectrum {\kosuke{agrees
    very well
    with the experimental phonon dispersion.}}
We found that alleviating the atomic basis-set redundancy of the trial wavefunction is key to reduce the statistical error of atomic forces and, thus, to make the VMC phonons calculations feasible over the full Brillouin zone.
This achievement paves the way to new relevant applications, 
{\kosuke{for instance,}}
in correlated materials {\kosuke{and Van der Waals crystals (i.e., molecular crystals~\cite{2011SUB, 2012CAS})}}, where sometimes phonons are poorly reproduced within the DFT framework.

\vspace{1mm}
\begin{acknowledgments}
K.N. is grateful for computational resources from the facilities of Research Center for Advanced Computing Infrastructure at Japan Advanced Institute of Science and Technology (JAIST).
K.N. and S.S. are grateful for computational resources from PRACE project No.~2019204934.
T.M. and M.C. acknowledge that this work was supported by French state funds managed by the ANR within the Investissements d'Avenir programme under reference  ANR-11-IDEX-0004-02, and more specifically within the framework of the Cluster of Excellence MATISSE led by Sorbonne University.
M.C. is grateful to the French Grand {\'e}quipement national de calcul intensif (GENCI) for the computational time provided through the Project No.~0906493. 
K.N., M.C., and S.S. are grateful for computational resources of the supercomputer Fugaku provided by RIKEN through the HPCI System Research Project (Project ID: hp200164).
{\kosuke{K.N. acknowledges a support from the JSPS Overseas Research Fellowships and that from Grant-in-Aid for Scientific Research on Innovative Areas (No.~16H06439).}}
R.M. is grateful for financial supports from 
MEXT-KAKENHI (19H04692 and 16KK0097), 
from Toyota Motor Corporation, 
from the Air Force Office of Scientific Research 
(AFOSR-AOARD/FA2386-17-1-4049;FA2386-19-1-4015), 
and from JSPS Bilateral Joint Projects (with India DST). 
S.S. also acknowledges a financial support from PRIN~2017BZPKSZ.
The authors appreciate helpful comments by A. Zen. 
\end{acknowledgments}

\appendix

\makeatletter
\renewcommand{\refname}{}
\renewcommand*{\citenumfont}[1]{#1}
\renewcommand*{\bibnumfmt}[1]{[#1]}
\makeatother

\setcounter{table}{0}
\setcounter{equation}{0}
\setcounter{figure}{0}
\renewcommand{\thetable}{S-\Roman{table}}
\renewcommand{\thefigure}{S-\arabic{figure}}
\renewcommand{\theequation}{S-\arabic{equation}}

\section{\label{qmc_details} Details of QMC calculations}

\vspace{3mm}
All Variational Monte Carlo (VMC) and lattice regularized diffusion Monte Carlo (LRDMC)~{\cite{2005CAS, 2020NAK1}} calculations in this study were performed by a SISSA quantum Monte Carlo package, called \tvb\ ~{\cite{2020NAK2}}. 
The package employs a many-body WF ansatz $\Psi$ which can be written as the product of two terms, i.e.,
$
\Psi  =  \Phi _\text{AS} \times \exp J \,,
$
where the term $\exp J$ and $\Phi _\text{AS}$ are conventionally called Jastrow and antisymmetric parts, respectively.
The antisymmetric part is denoted as the Antisymmetrized Geminal Power (AGP) that reads:
$
{\Psi _{{\text{AGP}}}}\left( {{{\mathbf{r}}_1}, \ldots ,{{\mathbf{r}}_N}} \right) = {\hat A} \left[ {\Phi \left( {{\mathbf{r}}_1^ \uparrow ,{\mathbf{r}}_1^ \downarrow } \right)\Phi \left( {{\mathbf{r}}_2^ \uparrow ,{\mathbf{r}}_2^ \downarrow } \right) \cdots \Phi \left( {{\mathbf{r}}_{N/2}^ \uparrow ,{\mathbf{r}}_{N/2}^ \downarrow } \right)} \right],
$
where ${\hat A}$ is the antisymmetrization operator, and $\Phi \left( {{\mathbf{r}}_{}^ \uparrow ,{\mathbf{r}}_{}^ \downarrow } \right)$ is called the paring function~\cite{2003CAS}. The spatial part of the geminal function is expanded over the Gaussian-type atomic orbitals:
$
{\Phi _{{\text{AGP}}}}\left( {{{\mathbf{r}}_i},{{\mathbf{r}}_j}} \right) = \sum\limits_{l,m,a,b} {{f_{\left\{ {a,l} \right\},\left\{ {b,m} \right\}}}{\psi _{a,l}}\left( {{{\mathbf{r}}_i}} \right){\psi _{b,m}}\left( {{{\mathbf{r}}_j}} \right)}
\label{agp_expansion}
$
where ${\psi _{a,l}}$ and ${\psi _{b,m}}$ are primitive Gaussian atomic orbitals, their indices $l$ and $m$ indicate different orbitals centered on atoms $a$ and $b$, and $i$ and $j$ are coordinates of spin up and down electrons, respectively, and ${{f_{\left\{ {a,l} \right\},\left\{ {b,m} \right\}}}}$ are the variational parameters.
In this study, a triple-zeta basis set (11$s$11$p$2$d$1$f$) accompanying an energy-consistent pseudo potential{~\cite{2007BUR}} was employed for the atomic orbitals of the antisymmetric part (Table~\ref{basis_set}).
The pairing function can be also written as
$
{\Phi _{{\text{AGPn}}}}\left( {{{\mathbf{r}}_i},{{\mathbf{r}}_j}} \right) = 
\sum_{k=1}^M \lambda_k \phi_k(\mathbf{r}_i) \phi_k(\mathbf{r}_j)
$
with $\lambda_{k} > 0$, where $\phi_k(\mathbf{r})$ is a molecular orbital, i.e., $\phi_k(\mathbf{r}) = \sum_{i=1}^{L}{c_{i,k}{\psi_{i}}(\mathbf{r})}$.
When the paring function is expanded over $M$ molecular orbitals where $M$ is equal to the half of the total number of electrons ($N/2$), the AGP coincides with the Slater-Determinant ansatz{~\cite{2017BEC, 2009MAR}}. In this study, we restricted ourselves to a Jastrow-Slater determinant (JSD) by setting $M = \frac{1}{2} \cdot N$, wherein the coefficients of atomic orbitals, i.e., ${c_{i,k}}$, were obtained by a Density Functional theory (DFT) calculation, and were fixed during a VMC optimization.

\begin{figure}[htbp]
  \centering
  \includegraphics[width=8.6cm]{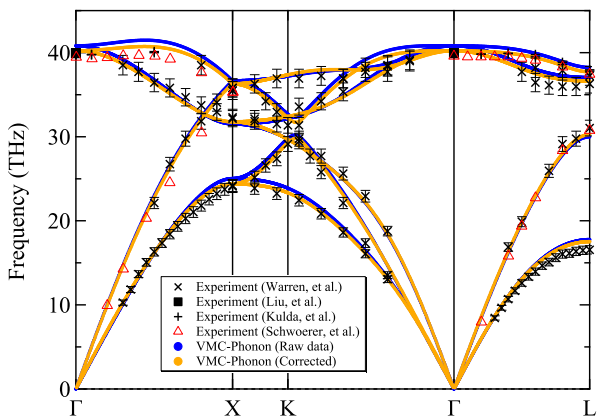}
  \caption{
  The phonon dispersions of diamond calculated using 2 $\times$ 2 $\times$ 2 conventional supercell at the VMC level. 
  ``Raw data" denotes the phonon dispersion obtained by the finite-displacement method, while
  ``Corrected" is the phonon dispersion where the anharmonic and one-body finite-size corrections are included.
  The error bars are not shown here for the sake of clarity. 
  {\it Observed} experimental results are also plotted for comparison. 
  Experimental data are taken from Refs.~{\onlinecite{1965WAR, *1967WAR, 2000LIU, 2002KUL, 1998SCH}}.   
  The data in Ref.~{\onlinecite{1998SCH}} were digitalized using WebPlotDigitizer~\cite{2020ROH}.
  }
  \label{pd_disp_diamond_diff}
\end{figure}

\begin{center}
\begin{table}[htbp]
\caption{\label{basis_set} The valence triple-zeta (VTZ) uncontracted basis set (11$s$11$p$2$d$1$f$) used for the determinant part in the present work.}

\begin{tabular}{c|c|c|c}
\Hline
Orbital type & Exponent & Orbital type & Exponent \\
\Hline
$s$ & 13.073594 & $p$ & 1.871016 \\
$s$ & 6.541187  & $p$ & 0.935757 \\
$s$ & 3.272791  & $p$ & 0.468003 \\
$s$ & 1.637494  & $p$ & 0.376742 \\
$s$ & 0.921552  & $p$ & 0.234064 \\
$s$ & 0.819297  & $p$ & 0.126772 \\
$s$ & 0.409924  & $p$ & 0.117063 \\
$s$ & 0.205100  & $p$ & 0.058547 \\
$s$ & 0.132800  & $p$ & 0.029281 \\
$s$ & 0.102619  & $d$ & 1.141611 \\
$s$ & 0.051344  & $d$ & 0.329486 \\
$p$ & 7.480076  & $f$ & 0.773485 \\
$p$ & 3.741035  &  &  \\
\Hline
\end{tabular}


\end{table}
\end{center}

%
\vspace{3mm}
The Jastrow term is composed of one-body, two-body and three/four-body factors ($J = {J_1}{J_2}{J_{3/4}}$). The one-body and two-body factors are essentially used to fulfill the electron-ion and electron-electron cusp conditions, respectively, and the three/four-body factor is employed to consider further electron-electron correlations (e.g., electron-nucleus-electron). Since we employed an energy-consistent pseudo potential developed by Burkatzki et al.~{\cite{2007BUR}}, we used only the inhomogeneous part of ${J_1^{inh}}$, $J_2$ and $J_{3/4}$ in this study (i.e., no electron-ion cusp corrections are needed). 
The inhomogeneous one-body Jastrow factor ${J_1^{inh}}$ reads
${J_1^{inh}}\left( {{{\mathbf{r}}_1}, \ldots, {{\mathbf{r}}_N}} \right) =  \sum_{i=1}^N \sum_{a=1}^{N_\text{atom}} \left( {\sum\limits_{l} {M_{a,l} \chi_{a,l}\left( {{{\mathbf{r}}_i}} \right)} } \right),
$
where ${{{\mathbf{r}}_i}}$ are the electron positions, ${{{\mathbf{R}}_a}}$ are the atomic positions with corresponding atomic number $Z_a$, $l$ runs over atomic orbitals $\chi _{a,l}$ ({\it e.g.}, GTO) centered on the atom $a$, ${N_\text{atom}}$ is the total number of atoms in a system, and
$\{ M_{a,l} \}$ are variational parameters.
The two-body Jastrow factor is defined as:
$
{J_2}\left( {{{\mathbf{r}}_1}, \ldots {{\mathbf{r}}_N}} \right) = \exp \left( {\sum\limits_{i < j} {v\left( {\left| {{{\mathbf{r}}_i} - {{\mathbf{r}}_j}} \right|} \right)} } \right),
$
where $v\left( r \right)$ is
$
v\left( r \right) = \frac{1}{2}r \cdot {\left( {1 - F \cdot r} \right)^{ - 1}}
$
and $F$ is a variational parameter. The three-body Jastrow factor is:
$
{J_{3/4}}\left( {{{\mathbf{r}}_1}, \ldots {{\mathbf{r}}_N}} \right) = \exp \left( {\sum\limits_{i < j} {{\Phi _{{\text{Jas}}}}\left( {{{\mathbf{r}}_i},{{\mathbf{r}}_j}} \right)} } \right),
$
and
$
{\Phi _{{\text{Jas}}}}\left( {{{\mathbf{r}}_i},{{\mathbf{r}}_j}} \right) = \sum\limits_{l,m,a,b} {g_{a,l,m,b}^{}\chi _{a,l}^{{\text{Jas}}}\left( {{{\mathbf{r}}_i}} \right)\chi _{b,m}^{{\text{Jas}}}\left( {{{\mathbf{r}}_j}} \right)},
$
where the indices $l$ and $m$ again indicate different orbitals centered on
corresponding atoms $a$ and $b$. In this study, the coefficients of the three/four-body Jastrow factor were set to zero for $a \neq b$ because it significantly decreases the number of variational parameters while rarely affects variational energies .
A basis set (3$s$2$p$) was employed for the atomic orbitals of the Jastrow part.
The variational parameters in the Jastrow factor were optimized by the so-called stochastic reconfiguration~\cite{2007SOR} and/or the modified linear method~\cite{2007UMR, 2020NAK2} implemented in \tvb.
Using the optimized wavefunction, energies and forces are calculated at the VMC and the LRDMC levels. All LRDMC calculations were performed by the original single-grid scheme~\cite{2005CAS} with the discretization grid size $a = 0.20$ Bohr.
For the Raman frequency calculation, several lattice spaces, i.e., $a$ = 0.20, 0.30, 0.35, 0.40, and 0.50 Bohr were also used to extrapolate the energies (i.e., $a \to 0$) with a quadratic function (i.e., $E(a) = E_0 + k_1 \cdot a^2 + k_2 \cdot a^4$).
%
%

%
\vspace{3mm}
Figure~\ref{pd_disp_diamond_diff} shows the phonon dispersions and Raman frequencies of diamond calculated using 2 $\times$ 2 $\times$ 2 conventional supercell at the VMC level. ``Raw data" denotes the phonon dispersion obtained by the finite-displacement method, while ``Corrected" is the phonon dispersion where the one-body finite-size (see sec.~{\ref{dft_validation}}) and anharmonic (see sec.~{\ref{pimd_renormalizations}}) corrections are included.

\begin{figure}[htbp]
  \centering
  \includegraphics[width=8.6cm]{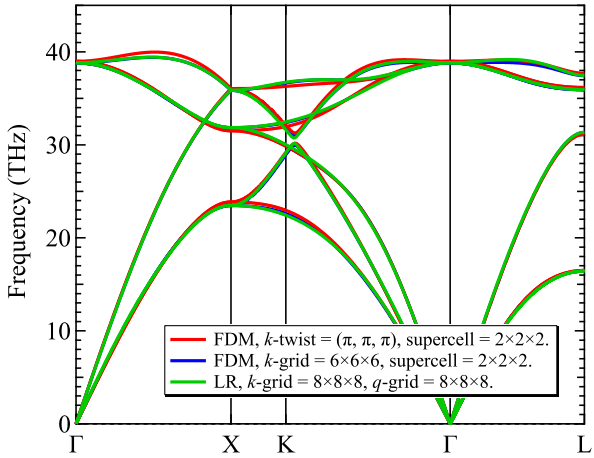}
  \caption{Phonon dispersions of diamond obtained by different methods with different computational conditions at the DFT level, wherein the experimental lattice parameter was used. (a) A phonon dispersion calculated by the finite-displacement method (FDM), or the so-called frozen-phonon method, implemented in Phonopy~\cite{2015TOG} with 2 $\times$ 2 $\times$ 2 conventional supercell at $L$ point. The displacement of 0.03 Bohr was employed for calculating derivatives of atomic forces with respect to a nucleus position. (b) The same as (b) except for $k$-grid. The shifted 6 $\times$ 6 $\times$ 6 Monkhorst-Pack grid~\cite{1976MON} was employed. (c) A phonon dispersion calculated by the linear response method implemented Quantum Espresso package~\cite{2009GIA} with 1 $\times$ 1 $\times$ 1 primitive unit cell. The shifted (unshifted) 8 $\times$ 8 $\times$ 8 Monkhorst-Pack grid~\cite{1976MON} was employed for $k$-integration ($q$-interpolation). The DFT calculations were performed with GGA-PBE functionals with Ultrasoft (US) pseudo potential taken from PS-Library~\cite{2014DAL}.
}
  \label{one-body-dft}
\end{figure}

\begin{figure}[htbp]
  \centering
  \includegraphics[width=8.6cm]{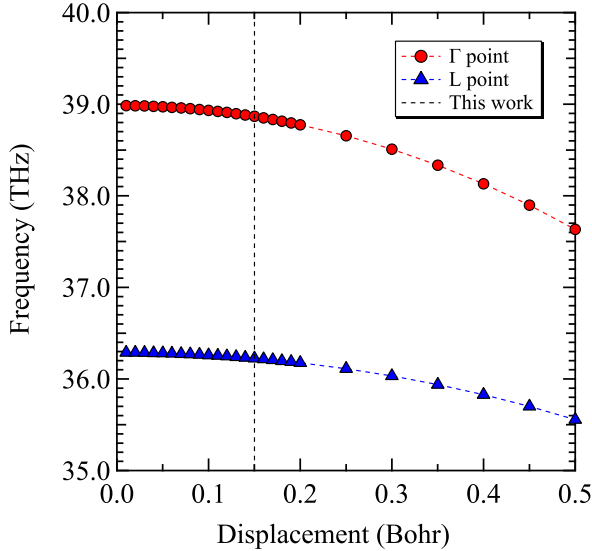}
  \caption{Phonon frequencies calculated by DFT at $q$ = $\Gamma$ and $L$ points v.s. employed displacements in the finite displacement method, where the GGA-PBE functional was employed. The phonon dispersions were calculated using 2 $\times$ 2 $\times$ 2 conventional supercell with $k$ twist = $L$ ($\pi$,$\pi$,$\pi$).}
  \label{fdm-dist-freq-conv}
\end{figure}

\begin{figure}[htbp]
  \centering
  \includegraphics[width=8.6cm]{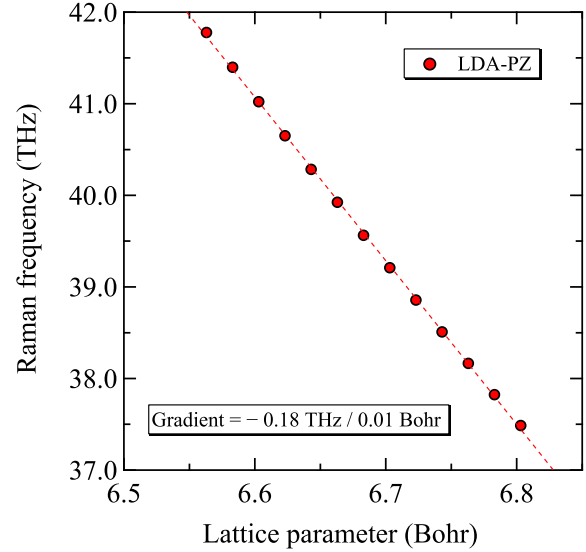}
  \caption{Plots of Raman Frequencies vs. lattice parameters of diamond.
They were calculated using the linear response theory implemented 
in Quantum Espresso at a single $q$ point, i.e., $q$ = (0,0,0) with
the LDA-PZ exchange-correlation functional. 
The primitive lattice was employed with $k$-grid = 8 $\times$ 8 $\times$ 8.
  }
  \label{Raman_freq_lattice_dep}
\end{figure}


\section{\label{dft_validation} Validation of computational conditions using DFT.}
There are two major implementations to compute a harmonic phonon dispersion, i.e., the finite displacement method (FDM) and the linear response method. The former is also called frozen phonon method. Their results should be consistent as far as employed hyperparameters are correct.
Figure~\ref{one-body-dft} shows phonon dispersions of diamond obtained by the two different methods at the DFT level, wherein the experimental lattice parameter was used. (a) A phonon dispersion calculated by the frozen-phonon method implemented in Phonopy~\cite{2015TOG} with 2 $\times$ 2 $\times$ 2 conventional supercell at $L$ point. The displacement of 0.03 Bohr was employed for calculating derivatives of atomic forces with respect to a nucleus position. (b) The same as (a) except for $k$-grid. The shifted 6 $\times$ 6 $\times$ 6 Monkhorst-Pack grid~\cite{1976MON} was employed. (c) A phonon dispersion calculated by the linear response method implemented Quantum Espresso package~\cite{2009GIA} with 1 $\times$ 1 $\times$ 1 primitive unit cell. The 8 $\times$ 8 $\times$ 8 Monkhorst-Pack grids~\cite{1976MON} were employed for $k$ and $q$ integrations. The DFT calculations were performed with GGA-PBE functionals with Ultrasoft (US) pseudo potential taken from PS-Library~\cite{2014DAL}.
Comparison of (a)-red with (c)-green proves that the one-body finite size error is very small in a phonon calculation thanks to the error cancellation when $L$ twist for 2 $\times$ 2 $\times$ 2 conventional supercell is employed. In detail, the one-body finite size errors are 0.16 THz, 0.18 THz, and 0.23 THz for $q$ = $\Gamma$, $X$, and $L$, respectively.
Comparison of (b)-blue with (c)-green, which are almost overlapped, suggests that the finite difference method does not bias the result. 
As mentioned in the main text, 0.15 Bohr was employed for the displacement of a carbon atom in the VMC frozen-phonon calculation to decrease the statistical error. Figure.~\ref{fdm-dist-freq-conv} shows the amplitude of 0.15 Bohr employed underestimates frequencies by $\sim$ 0.12~THz. A single-phonon phonon calculation at $\Gamma$ alleviates this underestimation as mentioned in the main text.
Fig.~{\ref{Raman_freq_lattice_dep}} shows a plot of Raman Frequencies vs. lattice parameters of diamond. They were calculated using the linear response theory implemented in Quantum Espresso at a single $q$ point, i.e., $q$ = (0,0,0) with the LDA-PZ exchange-correlation functional. The primitive lattice was employed with 8 $\times$ 8 $\times$ 8.

\section{\label{pimd_renormalizations} Anharmonic renormalizations from path integral molecular dynamics}
The anharmonic renormalizations of phonon frequencies reported in Table~\ref{Comp_ph_DFT_QMC} with the superscript '$f$' were computed through the displacement-displacement zero-time Kubo-correlator built upon path integral molecular dynamics (PIMD) trajectories. The evaluation of the displacement-displacement correlator over classical molecular dynamics trajectories is a standard method to predict force constant matrices and phonon spectra, but it misses nuclear quantum effects. The details of the extension of such a method to extract phonon dispersions from path integral molecular dynamics can be found in Ref.~\onlinecite{2020MOR}.

\vspace{3mm}
In Fig~\ref{pimd_spectrum}, we report the whole spectrum that we have obtained using this new approach compared to the linear response DFT curve.
\begin{figure}[htbp]
  \centering
  \includegraphics[width=8.6cm]{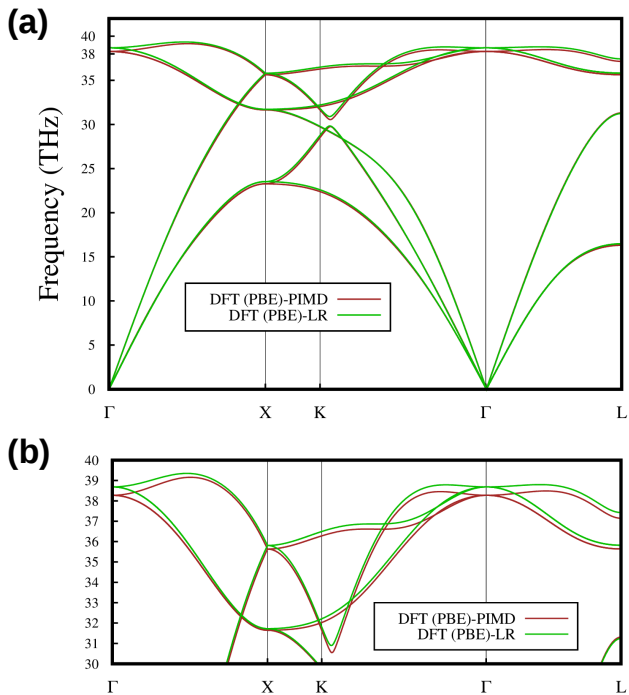}
  \caption{(a) Phonon spectrum at 300 K obtained from PIMD simulation
    versus linear response (LR) DFT spectrum; (b) Zoom over the optical modes.}
  \label{pimd_spectrum}
\end{figure}
In particular, the ion equations of motion were integrated using the PIOUD algorithm~\cite{2017MOU}, while at each time step the electronic potential energy surface was evaluated at DFT level using the Quantum Espresso package~\cite{2009GIA} and GGA-PBE~\cite{1996PER} exchange-correlation functionals. The simulation was carried out at 300 Kelvin for 34 picoseconds employing 12 beads. We used the same 64-atoms supercell as for the QMC calculations, with a 2 $\times$ 2 $\times$ 2 $k$-grid for electronic integration and a cutoff for wavefunctions equal to 60 Ry.

\begin{figure}[htbp]
  \centering
  \includegraphics[width=8.6cm]{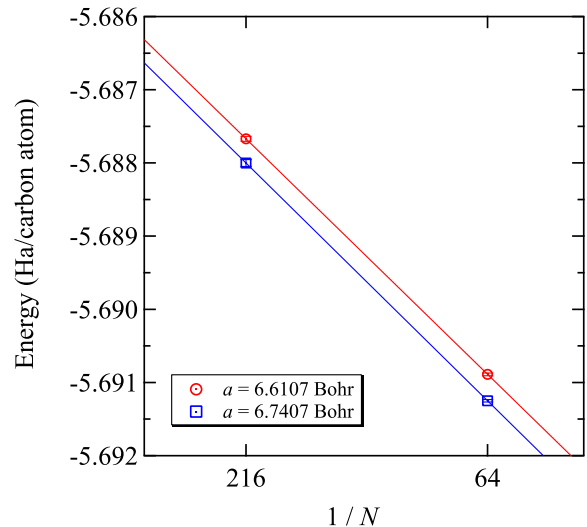}
  \caption{Finite-size extrapolation for different lattice parameters. The differences of energies are constant regardless of the simulation cell size, indicating that the finite-size error is negligible for an EOS calculation thanks to the error cancellation. Since diamond is an insulator, raw QMC energies obtained at $L$ point can be smoothly extrapolated by a linear function. See. Ref.{~\onlinecite{2010HEN}}
  }
  \label{finite-size-corr}
\end{figure}

\begin{figure}[htbp]
  \centering
  \includegraphics[width=8.4cm]{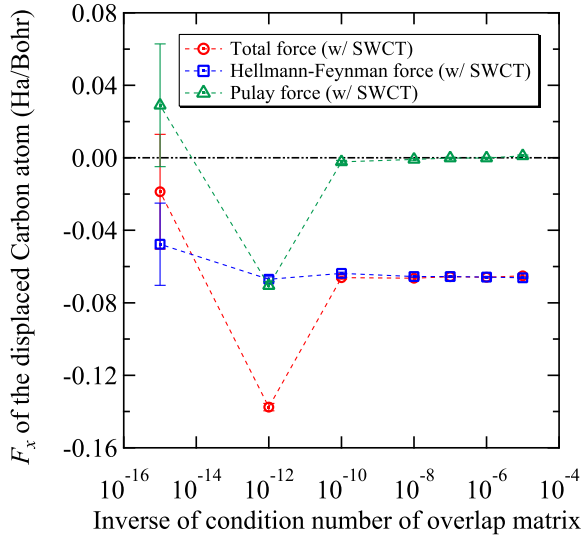}
  \caption{VMC forces, its Hellmann-Feynman, and its Pulay contributions v.s. inverse of the condition number of the overlap matrix.
The forces were calculated with the space warp coordinate transformations (SWCT)~\cite{2010SOR}. The same VTZ atomic basis set and the pseudo potential~{\cite{2007BUR}} as in the phonon and EOS calculations were adopted, while 1 $\times$ 1 $\times$ 1 conventional cell (8 carbon atoms in a simulation cell) with $k$ = $\Gamma$ twist was employed. Only one carbon atom was displaced in $x$ direction by 0.15 Bohr.
  }
  \label{err_force}
\end{figure}

\begin{center}
\begin{table*}[htbp]
\caption{\label{force_errors} Error bars of forces corresponding to Fig.~{\ref{epsover}}(b) in the main text. $\kappa ({\bf S})^{-1}$ denotes the inverse of the condition number of the overlap matrix ${\bf S}$. The used geometry is showin in Table~{\ref{displaced_diamond}}}

\begin{tabular}{c|cc|cc}
\Hline
\multirow{2}{*}{$\kappa ({\bf S})^{-1}$} & \multicolumn{2}{c|}{with SWCT} & \multicolumn{2}{c}{without SWCT} \\
\cline{2-5}
 & Force ($F_{x}$ of C1) (Ha/bohr) &  Error bar (Ha/bohr) & Force ($F_{x}$ of C1) (Ha/bohr) &  Error bar (Ha/bohr) \\
\Hline
$10^{-5}$  & -6.659 $\times$ $10^{-2}$  &  4.1 $\times$ $10^{-4}$  & -6.71 $\times$ $10^{-2}$ & 1.7 $\times$ $10^{-3}$ \\
$10^{-6}$  & -6.622 $\times$ $10^{-2}$  &  4.1 $\times$ $10^{-4}$  & -6.67 $\times$ $10^{-2}$ & 1.4 $\times$ $10^{-3}$ \\
$10^{-7}$  & -6.523 $\times$ $10^{-2}$  &  4.4 $\times$ $10^{-4}$  & -6.68 $\times$ $10^{-2}$ & 1.3 $\times$ $10^{-3}$  \\
$10^{-8}$  & -6.635 $\times$ $10^{-2}$  &  4.6 $\times$ $10^{-4}$  & -6.77 $\times$ $10^{-2}$ & 1.5 $\times$ $10^{-3}$  \\
$10^{-10}$  & -6.734 $\times$ $10^{-2}$  &  8.0 $\times$ $10^{-4}$ & -6.59 $\times$ $10^{-2}$ & 1.6 $\times$ $10^{-3}$ \\
$10^{-12}$  & -13.87 $\times$ $10^{-2}$  &  1.8 $\times$ $10^{-3}$ & -13.86 $\times$ $10^{-2}$ & 2.4 $\times$ $10^{-3}$  \\
$10^{-15}$  & -4.7   $\times$ $10^{-2}$  &  3.3 $\times$ $10^{-2}$ & -4.9   $\times$ $10^{-2}$ & 3.4 $\times$ $10^{-2}$ \\
\Hline

\end{tabular}


\end{table*}
\end{center}

\section{\label{fse_qmc} Finite-size errors in QMC calculations}

The effect of the finite-size errors on an EOS calculation has been investigated.
Figure~\ref{finite-size-corr} shows that finite-size extrapolations for two different lattice parameters ($a$ = 6.6107 and 6.7407 Bohr). The differences of the energies are constant regardless of the simulation cell size, indicating that the finite-size error is negligible for an EOS calculation thanks to the error cancellation. Since diamond is an insulator, raw QMC energies obtained at $L$ point can be smoothly extrapolated by a linear function. See. Ref.{~\onlinecite{2010HEN}}.

\section{\label{vinet_ap} Vinet exponential function}

In this study, the obtained energies were fitted by the Vinet exponential function~\cite{1987VIN} that reads:
\begin{eqnarray*}
E(V)&=&E_0 + \frac{2B_0V_0}{\left(B_0^\prime-1\right)^2}\left\lbrace 2 -\left[ 5 + 3\left( \frac{V}{V_0}\right)^\frac{1}{3} (B_0^\prime -1)  -3B_0^\prime \right] \right. \nonumber \\
&\times &\left. \exp \left[ -\frac{3}{2} \left( B_0^\prime-1\right)\left[\left( \frac{V}{V_0}\right)^\frac{1}{3} -1\right]\right]\right\rbrace,
\end{eqnarray*}
where $E(V)$ is the total energy per atom, $V$ is volume per atom, and $E_0$, $V_0$, $B_0$, and $B_0^\prime$ are parameters. The non-linear fittings were performed using SciPy module~\cite{2020VIR} implemented in Python.



%
\section{Atomic forces}
\tvb\ evaluates atomic forces in a differential expression{~\cite{2010SOR}}:
\begin{equation*}
{{\mathbf{F}}_a} =  - \left\langle {\frac{d}{{d{{\mathbf{R}}_a}}}{e_L}} \right\rangle  + 2\left( {\left\langle {{e_L}} \right\rangle \left\langle {\frac{d}{{d{{\mathbf{R}}_a}}}\ln \left( {{{\mathcal{J}}^{\frac{1}{2}}}\Psi } \right)} \right\rangle  - \left\langle {{e_L}\frac{d}{{d{{\mathbf{R}}_a}}}\ln \left( {{{\mathcal{J}}^{\frac{1}{2}}}\Psi } \right)} \right\rangle } \right)
\label{forces:aad}
\end{equation*}
where ${\mathcal{J}}$ is the Jacobian of the space warp coordinate transformation (SWCT){~\cite{2010SOR}}, ${e_L}$ is the local energy, and the brackets indicates a Monte Carlo average over the trial WF. 
The first term is called ``Hellmann-Feynman" force (HF) and the second and third terms are called ``Pulay" force (PF).
All the terms above can be written by the partial derivatives of the local energy and those of the logarithm of the WF{~\cite{2010SOR}}.
These differential expressions are efficiently computed in \tvb, by using the adjoint algorithmic differentiation technique{~\cite{2010SOR}}.
\tvb\ also employs the so-called reweighting technique developed by Attaccalite and Sorella~{\cite{2008ATT}} to avoid divergence of the derivatives in the vicinity of the nodal surfaces.
Fig.~{\ref{err_force}} shows VMC forces, its Hellmann-Feynman, and its Pulay contributions v.s. inverse of the condition number of the overlap matrix, corresponding to Fig.~\ref{epsover} in the main text. Fig.~{\ref{err_force}} clarifies that the erratic behavior of the forces at $\kappa ({\bf S})^{-1} = 10^{-12}$ comes from its Pulay contribution, indicating that the linear-dependency of a basis-set also deteriorates absolute values of forces.

\begin{center}
\begin{table}[htbp]
\caption{\label{displaced_diamond} The geometry of the displaced diamond used to obtain forces in Table{~\ref{force_errors}}. The lattice parameter is 6.741 Bohr. $x$, $y$, and $z$ are represented by fractional coordinates.}
\begin{tabular}{c|c|ccc}
\Hline
Element & Label & $x$ & $y$ & $z$ \\
\Hline
C & C1 &  0.02225 & 0.00000 & 0.00000 \\
C & C2 &  0.00000 & 0.50000 & 0.50000 \\
C & C3 &  0.50000 & 0.00000 & 0.50000 \\
C & C4 &  0.50000 & 0.50000 & 0.00000 \\
C & C5 &  0.25000 & 0.25000 & 0.75000 \\
C & C6 &  0.75000 & 0.75000 & 0.75000 \\
C & C7 &  0.25000 & 0.75000 & 0.25000 \\
C & C8 &  0.75000 & 0.25000 & 0.25000 \\
\Hline
\end{tabular}
\end{table}
\end{center}





\bibliographystyle{apsrev4-1}
\bibliography{./references.bib}

\begin{thebibliography}{69}%
\makeatletter
\providecommand \@ifxundefined [1]{%
 \@ifx{#1\undefined}
}%
\providecommand \@ifnum [1]{%
 \ifnum #1\expandafter \@firstoftwo
 \else \expandafter \@secondoftwo
 \fi
}%
\providecommand \@ifx [1]{%
 \ifx #1\expandafter \@firstoftwo
 \else \expandafter \@secondoftwo
 \fi
}%
\providecommand \natexlab [1]{#1}%
\providecommand \enquote  [1]{``#1''}%
\providecommand \bibnamefont  [1]{#1}%
\providecommand \bibfnamefont [1]{#1}%
\providecommand \citenamefont [1]{#1}%
\providecommand \href@noop [0]{\@secondoftwo}%
\providecommand \href [0]{\begingroup \@sanitize@url \@href}%
\providecommand \@href[1]{\@@startlink{#1}\@@href}%
\providecommand \@@href[1]{\endgroup#1\@@endlink}%
\providecommand \@sanitize@url [0]{\catcode `\\12\catcode `\$12\catcode
  `\&12\catcode `\#12\catcode `\^12\catcode `\_12\catcode `\%12\relax}%
\providecommand \@@startlink[1]{}%
\providecommand \@@endlink[0]{}%
\providecommand \url  [0]{\begingroup\@sanitize@url \@url }%
\providecommand \@url [1]{\endgroup\@href {#1}{\urlprefix }}%
\providecommand \urlprefix  [0]{URL }%
\providecommand \Eprint [0]{\href }%
\providecommand \doibase [0]{http://dx.doi.org/}%
\providecommand \selectlanguage [0]{\@gobble}%
\providecommand \bibinfo  [0]{\@secondoftwo}%
\providecommand \bibfield  [0]{\@secondoftwo}%
\providecommand \translation [1]{[#1]}%
\providecommand \BibitemOpen [0]{}%
\providecommand \bibitemStop [0]{}%
\providecommand \bibitemNoStop [0]{.\EOS\space}%
\providecommand \EOS [0]{\spacefactor3000\relax}%
\providecommand \BibitemShut  [1]{\csname bibitem#1\endcsname}%
\let\auto@bib@innerbib\@empty
\bibitem [{\citenamefont {Martin}(2004)}]{2004MAR}%
  \BibitemOpen
  \bibfield  {author} {\bibinfo {author} {\bibfnamefont {R.~M.}\ \bibnamefont
  {Martin}},\ }\href
  {https://books.google.it/books?hl=en&lr=&id=v1YhAwAAQBAJ&oi=fnd&pg=PR17&dq=martin+electronic+structure&ots=8Py35p2fcf&sig=m6qYd3VspZjCrjnB0olxmT4z3zQ#v=onepage&q=martin
  electronic structure&f=false} {\emph {\bibinfo {title} {{Electronic structure
  : basic theory and practical methods}}}}\ (\bibinfo  {publisher} {Cambridge
  University Press},\ \bibinfo {year} {2004})\BibitemShut {NoStop}%
\bibitem [{\citenamefont {Sholl}\ and\ \citenamefont
  {Steckel}(2011)}]{2011DAV}%
  \BibitemOpen
  \bibfield  {author} {\bibinfo {author} {\bibfnamefont {D.}~\bibnamefont
  {Sholl}}\ and\ \bibinfo {author} {\bibfnamefont {J.~A.}\ \bibnamefont
  {Steckel}},\ }\href@noop {} {\emph {\bibinfo {title} {Density functional
  theory: a practical introduction}}}\ (\bibinfo  {publisher} {John Wiley \&
  Sons},\ \bibinfo {year} {2011})\BibitemShut {NoStop}%
\bibitem [{\citenamefont {Baroni}\ \emph {et~al.}(2001)\citenamefont {Baroni},
  \citenamefont {de~Gironcoli}, \citenamefont {Dal~Corso},\ and\ \citenamefont
  {Giannozzi}}]{2001BAR}%
  \BibitemOpen
  \bibfield  {author} {\bibinfo {author} {\bibfnamefont {S.}~\bibnamefont
  {Baroni}}, \bibinfo {author} {\bibfnamefont {S.}~\bibnamefont
  {de~Gironcoli}}, \bibinfo {author} {\bibfnamefont {A.}~\bibnamefont
  {Dal~Corso}}, \ and\ \bibinfo {author} {\bibfnamefont {P.}~\bibnamefont
  {Giannozzi}},\ }\href {\doibase 10.1103/RevModPhys.73.515} {\bibfield
  {journal} {\bibinfo  {journal} {Rev. Mod. Phys.}\ }\textbf {\bibinfo {volume}
  {73}},\ \bibinfo {pages} {515} (\bibinfo {year} {2001})}\BibitemShut
  {NoStop}%
\bibitem [{\citenamefont {Togo}\ and\ \citenamefont {Tanaka}(2015)}]{2015TOG}%
  \BibitemOpen
  \bibfield  {author} {\bibinfo {author} {\bibfnamefont {A.}~\bibnamefont
  {Togo}}\ and\ \bibinfo {author} {\bibfnamefont {I.}~\bibnamefont {Tanaka}},\
  }\href@noop {} {\bibfield  {journal} {\bibinfo  {journal} {Scr. Mater.}\
  }\textbf {\bibinfo {volume} {108}},\ \bibinfo {pages} {1} (\bibinfo {year}
  {2015})}\BibitemShut {NoStop}%
\bibitem [{\citenamefont {Mounet}\ and\ \citenamefont
  {Marzari}(2005)}]{2005MOU}%
  \BibitemOpen
  \bibfield  {author} {\bibinfo {author} {\bibfnamefont {N.}~\bibnamefont
  {Mounet}}\ and\ \bibinfo {author} {\bibfnamefont {N.}~\bibnamefont
  {Marzari}},\ }\href@noop {} {\bibfield  {journal} {\bibinfo  {journal} {Phys.
  Rev. B}\ }\textbf {\bibinfo {volume} {71}},\ \bibinfo {pages} {205214}
  (\bibinfo {year} {2005})}\BibitemShut {NoStop}%
\bibitem [{\citenamefont {Lazzeri}\ \emph {et~al.}(2008)\citenamefont
  {Lazzeri}, \citenamefont {Attaccalite}, \citenamefont {Wirtz},\ and\
  \citenamefont {Mauri}}]{2008LAZ}%
  \BibitemOpen
  \bibfield  {author} {\bibinfo {author} {\bibfnamefont {M.}~\bibnamefont
  {Lazzeri}}, \bibinfo {author} {\bibfnamefont {C.}~\bibnamefont
  {Attaccalite}}, \bibinfo {author} {\bibfnamefont {L.}~\bibnamefont {Wirtz}},
  \ and\ \bibinfo {author} {\bibfnamefont {F.}~\bibnamefont {Mauri}},\
  }\href@noop {} {\bibfield  {journal} {\bibinfo  {journal} {Phys. Rev. B}\
  }\textbf {\bibinfo {volume} {78}},\ \bibinfo {pages} {081406(R)} (\bibinfo
  {year} {2008})}\BibitemShut {NoStop}%
\bibitem [{\citenamefont {Krisch}\ \emph {et~al.}(2011)\citenamefont {Krisch},
  \citenamefont {Farber}, \citenamefont {Xu}, \citenamefont {Antonangeli},
  \citenamefont {Aracne}, \citenamefont {Beraud}, \citenamefont {Chiang},
  \citenamefont {Zarestky}, \citenamefont {Kim}, \citenamefont {Isaev},
  \citenamefont {Ahuja},\ and\ \citenamefont {Johansson}}]{2011KRI}%
  \BibitemOpen
  \bibfield  {author} {\bibinfo {author} {\bibfnamefont {M.}~\bibnamefont
  {Krisch}}, \bibinfo {author} {\bibfnamefont {D.~L.}\ \bibnamefont {Farber}},
  \bibinfo {author} {\bibfnamefont {R.}~\bibnamefont {Xu}}, \bibinfo {author}
  {\bibfnamefont {D.}~\bibnamefont {Antonangeli}}, \bibinfo {author}
  {\bibfnamefont {C.~M.}\ \bibnamefont {Aracne}}, \bibinfo {author}
  {\bibfnamefont {A.}~\bibnamefont {Beraud}}, \bibinfo {author} {\bibfnamefont
  {T.-C.}\ \bibnamefont {Chiang}}, \bibinfo {author} {\bibfnamefont
  {J.}~\bibnamefont {Zarestky}}, \bibinfo {author} {\bibfnamefont {D.~Y.}\
  \bibnamefont {Kim}}, \bibinfo {author} {\bibfnamefont {E.~I.}\ \bibnamefont
  {Isaev}}, \bibinfo {author} {\bibfnamefont {R.}~\bibnamefont {Ahuja}}, \ and\
  \bibinfo {author} {\bibfnamefont {B.}~\bibnamefont {Johansson}},\ }\href
  {\doibase 10.1073/pnas.1015945108} {\bibfield  {journal} {\bibinfo  {journal}
  {Proc. Natl. Acad. Sci. U.S.A.}\ }\textbf {\bibinfo {volume} {108}},\
  \bibinfo {pages} {9342} (\bibinfo {year} {2011})}\BibitemShut {NoStop}%
\bibitem [{\citenamefont {Reznik}\ \emph {et~al.}(2008)\citenamefont {Reznik},
  \citenamefont {Sangiovanni}, \citenamefont {Gunnarsson},\ and\ \citenamefont
  {Devereaux}}]{2008REZ}%
  \BibitemOpen
  \bibfield  {author} {\bibinfo {author} {\bibfnamefont {D.}~\bibnamefont
  {Reznik}}, \bibinfo {author} {\bibfnamefont {G.}~\bibnamefont {Sangiovanni}},
  \bibinfo {author} {\bibfnamefont {O.}~\bibnamefont {Gunnarsson}}, \ and\
  \bibinfo {author} {\bibfnamefont {T.}~\bibnamefont {Devereaux}},\ }\href@noop
  {} {\bibfield  {journal} {\bibinfo  {journal} {Nature}\ }\textbf {\bibinfo
  {volume} {455}},\ \bibinfo {pages} {E6} (\bibinfo {year} {2008})}\BibitemShut
  {NoStop}%
\bibitem [{\citenamefont {Floris}\ \emph {et~al.}(2011)\citenamefont {Floris},
  \citenamefont {de~Gironcoli}, \citenamefont {Gross},\ and\ \citenamefont
  {Cococcioni}}]{2011FLO}%
  \BibitemOpen
  \bibfield  {author} {\bibinfo {author} {\bibfnamefont {A.}~\bibnamefont
  {Floris}}, \bibinfo {author} {\bibfnamefont {S.}~\bibnamefont
  {de~Gironcoli}}, \bibinfo {author} {\bibfnamefont {E.~K.~U.}\ \bibnamefont
  {Gross}}, \ and\ \bibinfo {author} {\bibfnamefont {M.}~\bibnamefont
  {Cococcioni}},\ }\href {\doibase 10.1103/PhysRevB.84.161102} {\bibfield
  {journal} {\bibinfo  {journal} {Phys. Rev. B}\ }\textbf {\bibinfo {volume}
  {84}},\ \bibinfo {pages} {161102(R)} (\bibinfo {year} {2011})}\BibitemShut
  {NoStop}%
\bibitem [{\citenamefont {Leonov}\ \emph {et~al.}(2012)\citenamefont {Leonov},
  \citenamefont {Poteryaev}, \citenamefont {Anisimov},\ and\ \citenamefont
  {Vollhardt}}]{2012LEO}%
  \BibitemOpen
  \bibfield  {author} {\bibinfo {author} {\bibfnamefont {I.}~\bibnamefont
  {Leonov}}, \bibinfo {author} {\bibfnamefont {A.~I.}\ \bibnamefont
  {Poteryaev}}, \bibinfo {author} {\bibfnamefont {V.~I.}\ \bibnamefont
  {Anisimov}}, \ and\ \bibinfo {author} {\bibfnamefont {D.}~\bibnamefont
  {Vollhardt}},\ }\href {\doibase 10.1103/PhysRevB.85.020401} {\bibfield
  {journal} {\bibinfo  {journal} {Phys. Rev. B}\ }\textbf {\bibinfo {volume}
  {85}},\ \bibinfo {pages} {020401(R)} (\bibinfo {year} {2012})}\BibitemShut
  {NoStop}%
\bibitem [{\citenamefont {Leonov}\ \emph {et~al.}(2014)\citenamefont {Leonov},
  \citenamefont {Anisimov},\ and\ \citenamefont {Vollhardt}}]{2014LEO}%
  \BibitemOpen
  \bibfield  {author} {\bibinfo {author} {\bibfnamefont {I.}~\bibnamefont
  {Leonov}}, \bibinfo {author} {\bibfnamefont {V.~I.}\ \bibnamefont
  {Anisimov}}, \ and\ \bibinfo {author} {\bibfnamefont {D.}~\bibnamefont
  {Vollhardt}},\ }\href {\doibase 10.1103/PhysRevLett.112.146401} {\bibfield
  {journal} {\bibinfo  {journal} {Phys. Rev. Lett.}\ }\textbf {\bibinfo
  {volume} {112}},\ \bibinfo {pages} {146401} (\bibinfo {year}
  {2014})}\BibitemShut {NoStop}%
\bibitem [{\citenamefont {Foulkes}\ \emph {et~al.}(2001)\citenamefont
  {Foulkes}, \citenamefont {Mitas}, \citenamefont {Needs},\ and\ \citenamefont
  {Rajagopal}}]{2001FOU}%
  \BibitemOpen
  \bibfield  {author} {\bibinfo {author} {\bibfnamefont {W.}~\bibnamefont
  {Foulkes}}, \bibinfo {author} {\bibfnamefont {L.}~\bibnamefont {Mitas}},
  \bibinfo {author} {\bibfnamefont {R.}~\bibnamefont {Needs}}, \ and\ \bibinfo
  {author} {\bibfnamefont {G.}~\bibnamefont {Rajagopal}},\ }\href@noop {}
  {\bibfield  {journal} {\bibinfo  {journal} {Rev. Mod. Phys.}\ }\textbf
  {\bibinfo {volume} {73}},\ \bibinfo {pages} {33} (\bibinfo {year}
  {2001})}\BibitemShut {NoStop}%
\bibitem [{\citenamefont {Marchi}\ \emph {et~al.}(2011)\citenamefont {Marchi},
  \citenamefont {Azadi},\ and\ \citenamefont {Sorella}}]{2011MAR}%
  \BibitemOpen
  \bibfield  {author} {\bibinfo {author} {\bibfnamefont {M.}~\bibnamefont
  {Marchi}}, \bibinfo {author} {\bibfnamefont {S.}~\bibnamefont {Azadi}}, \
  and\ \bibinfo {author} {\bibfnamefont {S.}~\bibnamefont {Sorella}},\
  }\href@noop {} {\bibfield  {journal} {\bibinfo  {journal} {Phys. Rev. Lett.}\
  }\textbf {\bibinfo {volume} {107}},\ \bibinfo {pages} {086807} (\bibinfo
  {year} {2011})}\BibitemShut {NoStop}%
\bibitem [{\citenamefont {Casula}\ and\ \citenamefont
  {Sorella}(2013)}]{2013CAS}%
  \BibitemOpen
  \bibfield  {author} {\bibinfo {author} {\bibfnamefont {M.}~\bibnamefont
  {Casula}}\ and\ \bibinfo {author} {\bibfnamefont {S.}~\bibnamefont
  {Sorella}},\ }\href@noop {} {\bibfield  {journal} {\bibinfo  {journal} {Phys.
  Rev. B}\ }\textbf {\bibinfo {volume} {88}},\ \bibinfo {pages} {155125}
  (\bibinfo {year} {2013})}\BibitemShut {NoStop}%
\bibitem [{\citenamefont {Sorella}\ \emph {et~al.}(2018)\citenamefont
  {Sorella}, \citenamefont {Seki}, \citenamefont {Brovko}, \citenamefont
  {Shirakawa}, \citenamefont {Miyakoshi}, \citenamefont {Yunoki},\ and\
  \citenamefont {Tosatti}}]{2018SOR}%
  \BibitemOpen
  \bibfield  {author} {\bibinfo {author} {\bibfnamefont {S.}~\bibnamefont
  {Sorella}}, \bibinfo {author} {\bibfnamefont {K.}~\bibnamefont {Seki}},
  \bibinfo {author} {\bibfnamefont {O.~O.}\ \bibnamefont {Brovko}}, \bibinfo
  {author} {\bibfnamefont {T.}~\bibnamefont {Shirakawa}}, \bibinfo {author}
  {\bibfnamefont {S.}~\bibnamefont {Miyakoshi}}, \bibinfo {author}
  {\bibfnamefont {S.}~\bibnamefont {Yunoki}}, \ and\ \bibinfo {author}
  {\bibfnamefont {E.}~\bibnamefont {Tosatti}},\ }\href@noop {} {\bibfield
  {journal} {\bibinfo  {journal} {Phys. Rev. Lett.}\ }\textbf {\bibinfo
  {volume} {121}},\ \bibinfo {pages} {066402} (\bibinfo {year}
  {2018})}\BibitemShut {NoStop}%
\bibitem [{\citenamefont {Zen}\ \emph {et~al.}(2012)\citenamefont {Zen},
  \citenamefont {Zhelyazov},\ and\ \citenamefont {Guidoni}}]{2012ZEN}%
  \BibitemOpen
  \bibfield  {author} {\bibinfo {author} {\bibfnamefont {A.}~\bibnamefont
  {Zen}}, \bibinfo {author} {\bibfnamefont {D.}~\bibnamefont {Zhelyazov}}, \
  and\ \bibinfo {author} {\bibfnamefont {L.}~\bibnamefont {Guidoni}},\
  }\href@noop {} {\bibfield  {journal} {\bibinfo  {journal} {J. Chem. Theory
  Comput.}\ }\textbf {\bibinfo {volume} {8}},\ \bibinfo {pages} {4204}
  (\bibinfo {year} {2012})}\BibitemShut {NoStop}%
\bibitem [{\citenamefont {Luo}\ \emph {et~al.}(2014)\citenamefont {Luo},
  \citenamefont {Zen},\ and\ \citenamefont {Sorella}}]{2014LUO}%
  \BibitemOpen
  \bibfield  {author} {\bibinfo {author} {\bibfnamefont {Y.}~\bibnamefont
  {Luo}}, \bibinfo {author} {\bibfnamefont {A.}~\bibnamefont {Zen}}, \ and\
  \bibinfo {author} {\bibfnamefont {S.}~\bibnamefont {Sorella}},\ }\href@noop
  {} {\bibfield  {journal} {\bibinfo  {journal} {J. Chem. Phys.}\ }\textbf
  {\bibinfo {volume} {141}},\ \bibinfo {pages} {194112} (\bibinfo {year}
  {2014})}\BibitemShut {NoStop}%
\bibitem [{\citenamefont {Liu}\ \emph {et~al.}(2019)\citenamefont {Liu},
  \citenamefont {Andrews},\ and\ \citenamefont {Conduit}}]{2019LIU}%
  \BibitemOpen
  \bibfield  {author} {\bibinfo {author} {\bibfnamefont {Y.~Y.~F.}\
  \bibnamefont {Liu}}, \bibinfo {author} {\bibfnamefont {B.}~\bibnamefont
  {Andrews}}, \ and\ \bibinfo {author} {\bibfnamefont {G.~J.}\ \bibnamefont
  {Conduit}},\ }\href@noop {} {\bibfield  {journal} {\bibinfo  {journal} {J.
  Chem. Phys.}\ }\textbf {\bibinfo {volume} {150}},\ \bibinfo {pages} {034104}
  (\bibinfo {year} {2019})}\BibitemShut {NoStop}%
\bibitem [{\citenamefont {Nakano}\ \emph {et~al.}(2019)\citenamefont {Nakano},
  \citenamefont {Maezono},\ and\ \citenamefont {Sorella}}]{2019NAK}%
  \BibitemOpen
  \bibfield  {author} {\bibinfo {author} {\bibfnamefont {K.}~\bibnamefont
  {Nakano}}, \bibinfo {author} {\bibfnamefont {R.}~\bibnamefont {Maezono}}, \
  and\ \bibinfo {author} {\bibfnamefont {S.}~\bibnamefont {Sorella}},\ }\href
  {\doibase 10.1021/acs.jctc.9b00295} {\bibfield  {journal} {\bibinfo
  {journal} {J. Chem. Theory Comput.}\ }\textbf {\bibinfo {volume} {15}},\
  \bibinfo {pages} {4044} (\bibinfo {year} {2019})}\BibitemShut {NoStop}%
\bibitem [{\citenamefont {Maezono}\ \emph {et~al.}(2007)\citenamefont
  {Maezono}, \citenamefont {Ma}, \citenamefont {Towler},\ and\ \citenamefont
  {Needs}}]{2007MAE}%
  \BibitemOpen
  \bibfield  {author} {\bibinfo {author} {\bibfnamefont {R.}~\bibnamefont
  {Maezono}}, \bibinfo {author} {\bibfnamefont {A.}~\bibnamefont {Ma}},
  \bibinfo {author} {\bibfnamefont {M.~D.}\ \bibnamefont {Towler}}, \ and\
  \bibinfo {author} {\bibfnamefont {R.~J.}\ \bibnamefont {Needs}},\ }\href
  {\doibase 10.1103/PhysRevLett.98.025701} {\bibfield  {journal} {\bibinfo
  {journal} {Phys. Rev. Lett.}\ }\textbf {\bibinfo {volume} {98}},\ \bibinfo
  {pages} {025701} (\bibinfo {year} {2007})}\BibitemShut {NoStop}%
\bibitem [{\citenamefont {Azadi}\ \emph {et~al.}(2010)\citenamefont {Azadi},
  \citenamefont {Cavazzoni},\ and\ \citenamefont {Sorella}}]{2010AZA}%
  \BibitemOpen
  \bibfield  {author} {\bibinfo {author} {\bibfnamefont {S.}~\bibnamefont
  {Azadi}}, \bibinfo {author} {\bibfnamefont {C.}~\bibnamefont {Cavazzoni}}, \
  and\ \bibinfo {author} {\bibfnamefont {S.}~\bibnamefont {Sorella}},\
  }\href@noop {} {\bibfield  {journal} {\bibinfo  {journal} {Phys. Rev. B}\
  }\textbf {\bibinfo {volume} {82}},\ \bibinfo {pages} {125112} (\bibinfo
  {year} {2010})}\BibitemShut {NoStop}%
\bibitem [{\citenamefont {Liu}\ \emph {et~al.}(2000)\citenamefont {Liu},
  \citenamefont {Bursill}, \citenamefont {Prawer},\ and\ \citenamefont
  {Beserman}}]{2000LIU}%
  \BibitemOpen
  \bibfield  {author} {\bibinfo {author} {\bibfnamefont {M.~S.}\ \bibnamefont
  {Liu}}, \bibinfo {author} {\bibfnamefont {L.~A.}\ \bibnamefont {Bursill}},
  \bibinfo {author} {\bibfnamefont {S.}~\bibnamefont {Prawer}}, \ and\ \bibinfo
  {author} {\bibfnamefont {R.}~\bibnamefont {Beserman}},\ }\href {\doibase
  10.1103/PhysRevB.61.3391} {\bibfield  {journal} {\bibinfo  {journal} {Phys.
  Rev. B}\ }\textbf {\bibinfo {volume} {61}},\ \bibinfo {pages} {3391}
  (\bibinfo {year} {2000})}\BibitemShut {NoStop}%
\bibitem [{\citenamefont {Schwoerer-B\"ohning}\ \emph
  {et~al.}(1998)\citenamefont {Schwoerer-B\"ohning}, \citenamefont
  {Macrander},\ and\ \citenamefont {Arms}}]{1998SCH}%
  \BibitemOpen
  \bibfield  {author} {\bibinfo {author} {\bibfnamefont {M.}~\bibnamefont
  {Schwoerer-B\"ohning}}, \bibinfo {author} {\bibfnamefont {A.~T.}\
  \bibnamefont {Macrander}}, \ and\ \bibinfo {author} {\bibfnamefont {D.~A.}\
  \bibnamefont {Arms}},\ }\href {\doibase 10.1103/PhysRevLett.80.5572}
  {\bibfield  {journal} {\bibinfo  {journal} {Phys. Rev. Lett.}\ }\textbf
  {\bibinfo {volume} {80}},\ \bibinfo {pages} {5572} (\bibinfo {year}
  {1998})}\BibitemShut {NoStop}%
\bibitem [{\citenamefont {Kulda}\ \emph {et~al.}(2002)\citenamefont {Kulda},
  \citenamefont {Kainzmaier}, \citenamefont {Strauch}, \citenamefont {Dorner},
  \citenamefont {Lorenzen},\ and\ \citenamefont {Krisch}}]{2002KUL}%
  \BibitemOpen
  \bibfield  {author} {\bibinfo {author} {\bibfnamefont {J.}~\bibnamefont
  {Kulda}}, \bibinfo {author} {\bibfnamefont {H.}~\bibnamefont {Kainzmaier}},
  \bibinfo {author} {\bibfnamefont {D.}~\bibnamefont {Strauch}}, \bibinfo
  {author} {\bibfnamefont {B.}~\bibnamefont {Dorner}}, \bibinfo {author}
  {\bibfnamefont {M.}~\bibnamefont {Lorenzen}}, \ and\ \bibinfo {author}
  {\bibfnamefont {M.}~\bibnamefont {Krisch}},\ }\href {\doibase
  10.1103/PhysRevB.66.241202} {\bibfield  {journal} {\bibinfo  {journal} {Phys.
  Rev. B}\ }\textbf {\bibinfo {volume} {66}},\ \bibinfo {pages} {241202(R)}
  (\bibinfo {year} {2002})}\BibitemShut {NoStop}%
\bibitem [{\citenamefont {Vanderbilt}\ \emph {et~al.}(1984)\citenamefont
  {Vanderbilt}, \citenamefont {Louie},\ and\ \citenamefont {Cohen}}]{1984VAN}%
  \BibitemOpen
  \bibfield  {author} {\bibinfo {author} {\bibfnamefont {D.}~\bibnamefont
  {Vanderbilt}}, \bibinfo {author} {\bibfnamefont {S.~G.}\ \bibnamefont
  {Louie}}, \ and\ \bibinfo {author} {\bibfnamefont {M.~L.}\ \bibnamefont
  {Cohen}},\ }\href {\doibase 10.1103/PhysRevLett.53.1477} {\bibfield
  {journal} {\bibinfo  {journal} {Phys. Rev. Lett.}\ }\textbf {\bibinfo
  {volume} {53}},\ \bibinfo {pages} {1477} (\bibinfo {year}
  {1984})}\BibitemShut {NoStop}%
\bibitem [{\citenamefont {Parrish}(1960)}]{1960PAR}%
  \BibitemOpen
  \bibfield  {author} {\bibinfo {author} {\bibfnamefont {W.}~\bibnamefont
  {Parrish}},\ }\href@noop {} {\bibfield  {journal} {\bibinfo  {journal} {Acta
  Cryst.}\ }\textbf {\bibinfo {volume} {13}},\ \bibinfo {pages} {838} (\bibinfo
  {year} {1960})}\BibitemShut {NoStop}%
\bibitem [{\citenamefont {Warren}\ \emph {et~al.}(1965)\citenamefont {Warren},
  \citenamefont {Wenzel},\ and\ \citenamefont {Yarnell}}]{1965WAR}%
  \BibitemOpen
  \bibfield  {author} {\bibinfo {author} {\bibfnamefont {J.~L.}\ \bibnamefont
  {Warren}}, \bibinfo {author} {\bibfnamefont {R.~G.}\ \bibnamefont {Wenzel}},
  \ and\ \bibinfo {author} {\bibfnamefont {J.~L.}\ \bibnamefont {Yarnell}},\
  }in\ \href@noop {} {\emph {\bibinfo {booktitle} {Inelastic Scattering of
  Neutrons}}},\ Vol.~\bibinfo {volume} {1}\ (\bibinfo  {publisher}
  {International Atomic Energy Agency},\ \bibinfo {address} {Vienna},\ \bibinfo
  {year} {1965})\ pp.\ \bibinfo {pages} {361--371}\BibitemShut {NoStop}%
\bibitem [{\citenamefont {Warren}\ \emph {et~al.}(1967)\citenamefont {Warren},
  \citenamefont {Yarnell}, \citenamefont {Dolling},\ and\ \citenamefont
  {Cowley}}]{1967WAR}%
  \BibitemOpen
  \bibfield  {author} {\bibinfo {author} {\bibfnamefont {J.~L.}\ \bibnamefont
  {Warren}}, \bibinfo {author} {\bibfnamefont {J.~L.}\ \bibnamefont {Yarnell}},
  \bibinfo {author} {\bibfnamefont {G.}~\bibnamefont {Dolling}}, \ and\
  \bibinfo {author} {\bibfnamefont {R.~A.}\ \bibnamefont {Cowley}},\ }\href
  {\doibase 10.1103/PhysRev.158.805} {\bibfield  {journal} {\bibinfo  {journal}
  {Phys. Rev.}\ }\textbf {\bibinfo {volume} {158}},\ \bibinfo {pages} {805}
  (\bibinfo {year} {1967})}\BibitemShut {NoStop}%
\bibitem [{\citenamefont {Rohatgi}(2020)}]{2020ROH}%
  \BibitemOpen
  \bibfield  {author} {\bibinfo {author} {\bibfnamefont {A.}~\bibnamefont
  {Rohatgi}},\ }\href@noop {} {\enquote {\bibinfo {title} {Webplotdigitizer:
  Version 4.3},}\ }\bibinfo {howpublished}
  {\url{https://automeris.io/WebPlotDigitizer}} (\bibinfo {year}
  {2020})\BibitemShut {NoStop}%
\bibitem [{\citenamefont {Casula}\ \emph {et~al.}(2005)\citenamefont {Casula},
  \citenamefont {Filippi},\ and\ \citenamefont {Sorella}}]{2005CAS}%
  \BibitemOpen
  \bibfield  {author} {\bibinfo {author} {\bibfnamefont {M.}~\bibnamefont
  {Casula}}, \bibinfo {author} {\bibfnamefont {C.}~\bibnamefont {Filippi}}, \
  and\ \bibinfo {author} {\bibfnamefont {S.}~\bibnamefont {Sorella}},\
  }\href@noop {} {\bibfield  {journal} {\bibinfo  {journal} {Phys. Rev. Lett.}\
  }\textbf {\bibinfo {volume} {95}},\ \bibinfo {pages} {100201} (\bibinfo
  {year} {2005})}\BibitemShut {NoStop}%
\bibitem [{\citenamefont {Nakano}\ \emph
  {et~al.}(2020{\natexlab{a}})\citenamefont {Nakano}, \citenamefont {Maezono},\
  and\ \citenamefont {Sorella}}]{2020NAK1}%
  \BibitemOpen
  \bibfield  {author} {\bibinfo {author} {\bibfnamefont {K.}~\bibnamefont
  {Nakano}}, \bibinfo {author} {\bibfnamefont {R.}~\bibnamefont {Maezono}}, \
  and\ \bibinfo {author} {\bibfnamefont {S.}~\bibnamefont {Sorella}},\ }\href
  {\doibase 10.1103/PhysRevB.101.155106} {\bibfield  {journal} {\bibinfo
  {journal} {Phys. Rev. B}\ }\textbf {\bibinfo {volume} {101}},\ \bibinfo
  {pages} {155106} (\bibinfo {year} {2020}{\natexlab{a}})}\BibitemShut
  {NoStop}%
\bibitem [{\citenamefont {Casula}\ and\ \citenamefont
  {Sorella}(2003)}]{2003CAS}%
  \BibitemOpen
  \bibfield  {author} {\bibinfo {author} {\bibfnamefont {M.}~\bibnamefont
  {Casula}}\ and\ \bibinfo {author} {\bibfnamefont {S.}~\bibnamefont
  {Sorella}},\ }\href@noop {} {\bibfield  {journal} {\bibinfo  {journal} {J.
  Chem. Phys.}\ }\textbf {\bibinfo {volume} {119}},\ \bibinfo {pages} {6500}
  (\bibinfo {year} {2003})}\BibitemShut {NoStop}%
\bibitem [{\citenamefont {Nakano}\ \emph
  {et~al.}(2020{\natexlab{b}})\citenamefont {Nakano}, \citenamefont
  {Attaccalite}, \citenamefont {Barborini}, \citenamefont {Capriotti},
  \citenamefont {Casula}, \citenamefont {Coccia}, \citenamefont {Dagrada},
  \citenamefont {Genovese}, \citenamefont {Luo}, \citenamefont {Mazzola} \emph
  {et~al.}}]{2020NAK2}%
  \BibitemOpen
  \bibfield  {author} {\bibinfo {author} {\bibfnamefont {K.}~\bibnamefont
  {Nakano}}, \bibinfo {author} {\bibfnamefont {C.}~\bibnamefont {Attaccalite}},
  \bibinfo {author} {\bibfnamefont {M.}~\bibnamefont {Barborini}}, \bibinfo
  {author} {\bibfnamefont {L.}~\bibnamefont {Capriotti}}, \bibinfo {author}
  {\bibfnamefont {M.}~\bibnamefont {Casula}}, \bibinfo {author} {\bibfnamefont
  {E.}~\bibnamefont {Coccia}}, \bibinfo {author} {\bibfnamefont
  {M.}~\bibnamefont {Dagrada}}, \bibinfo {author} {\bibfnamefont
  {C.}~\bibnamefont {Genovese}}, \bibinfo {author} {\bibfnamefont
  {Y.}~\bibnamefont {Luo}}, \bibinfo {author} {\bibfnamefont {G.}~\bibnamefont
  {Mazzola}},  \emph {et~al.},\ }\href@noop {} {\bibfield  {journal} {\bibinfo
  {journal} {J. Chem. Phys.}\ }\textbf {\bibinfo {volume} {152}},\ \bibinfo
  {pages} {204121} (\bibinfo {year} {2020}{\natexlab{b}})}\BibitemShut
  {NoStop}%
\bibitem [{\citenamefont {Burkatzki}\ \emph {et~al.}(2007)\citenamefont
  {Burkatzki}, \citenamefont {Filippi},\ and\ \citenamefont {Dolg}}]{2007BUR}%
  \BibitemOpen
  \bibfield  {author} {\bibinfo {author} {\bibfnamefont {M.}~\bibnamefont
  {Burkatzki}}, \bibinfo {author} {\bibfnamefont {C.}~\bibnamefont {Filippi}},
  \ and\ \bibinfo {author} {\bibfnamefont {M.}~\bibnamefont {Dolg}},\
  }\href@noop {} {\bibfield  {journal} {\bibinfo  {journal} {J. Chem. Phys.}\
  }\textbf {\bibinfo {volume} {126}},\ \bibinfo {pages} {234105} (\bibinfo
  {year} {2007})}\BibitemShut {NoStop}%
\bibitem [{foo()}]{foot}%
  \BibitemOpen
  \href@noop {} {}\bibinfo {howpublished} {See Supplemental Material for
  details of the DFT and QMC calculations, and for the supplemental Figures
  S-1-S-7 and Tables S-I-S-III.}\BibitemShut {Stop}%
\bibitem [{\citenamefont {Perdew}\ and\ \citenamefont
  {Zunger}(1981)}]{1981PER}%
  \BibitemOpen
  \bibfield  {author} {\bibinfo {author} {\bibfnamefont {J.~P.}\ \bibnamefont
  {Perdew}}\ and\ \bibinfo {author} {\bibfnamefont {A.}~\bibnamefont
  {Zunger}},\ }\href {\doibase 10.1103/PhysRevB.23.5048} {\bibfield  {journal}
  {\bibinfo  {journal} {Phys. Rev. B}\ }\textbf {\bibinfo {volume} {23}},\
  \bibinfo {pages} {5048} (\bibinfo {year} {1981})}\BibitemShut {NoStop}%
\bibitem [{\citenamefont {Sorella}\ \emph {et~al.}(2007)\citenamefont
  {Sorella}, \citenamefont {Casula},\ and\ \citenamefont {Rocca}}]{2007SOR}%
  \BibitemOpen
  \bibfield  {author} {\bibinfo {author} {\bibfnamefont {S.}~\bibnamefont
  {Sorella}}, \bibinfo {author} {\bibfnamefont {M.}~\bibnamefont {Casula}}, \
  and\ \bibinfo {author} {\bibfnamefont {D.}~\bibnamefont {Rocca}},\
  }\href@noop {} {\bibfield  {journal} {\bibinfo  {journal} {J. Chem. Phys.}\
  }\textbf {\bibinfo {volume} {127}},\ \bibinfo {pages} {014105} (\bibinfo
  {year} {2007})}\BibitemShut {NoStop}%
\bibitem [{\citenamefont {Umrigar}\ \emph {et~al.}(2007)\citenamefont
  {Umrigar}, \citenamefont {Toulouse}, \citenamefont {Filippi}, \citenamefont
  {Sorella},\ and\ \citenamefont {Hennig}}]{2007UMR}%
  \BibitemOpen
  \bibfield  {author} {\bibinfo {author} {\bibfnamefont {C.~J.}\ \bibnamefont
  {Umrigar}}, \bibinfo {author} {\bibfnamefont {J.}~\bibnamefont {Toulouse}},
  \bibinfo {author} {\bibfnamefont {C.}~\bibnamefont {Filippi}}, \bibinfo
  {author} {\bibfnamefont {S.}~\bibnamefont {Sorella}}, \ and\ \bibinfo
  {author} {\bibfnamefont {R.~G.}\ \bibnamefont {Hennig}},\ }\href {\doibase
  10.1103/PhysRevLett.98.110201} {\bibfield  {journal} {\bibinfo  {journal}
  {Phys. Rev. Lett.}\ }\textbf {\bibinfo {volume} {98}},\ \bibinfo {pages}
  {110201} (\bibinfo {year} {2007})}\BibitemShut {NoStop}%
\bibitem [{\citenamefont {Hennig}\ \emph {et~al.}(2010)\citenamefont {Hennig},
  \citenamefont {Wadehra}, \citenamefont {Driver}, \citenamefont {Parker},
  \citenamefont {Umrigar},\ and\ \citenamefont {Wilkins}}]{2010HEN}%
  \BibitemOpen
  \bibfield  {author} {\bibinfo {author} {\bibfnamefont {R.~G.}\ \bibnamefont
  {Hennig}}, \bibinfo {author} {\bibfnamefont {A.}~\bibnamefont {Wadehra}},
  \bibinfo {author} {\bibfnamefont {K.~P.}\ \bibnamefont {Driver}}, \bibinfo
  {author} {\bibfnamefont {W.~D.}\ \bibnamefont {Parker}}, \bibinfo {author}
  {\bibfnamefont {C.~J.}\ \bibnamefont {Umrigar}}, \ and\ \bibinfo {author}
  {\bibfnamefont {J.~W.}\ \bibnamefont {Wilkins}},\ }\href {\doibase
  10.1103/PhysRevB.82.014101} {\bibfield  {journal} {\bibinfo  {journal} {Phys.
  Rev. B}\ }\textbf {\bibinfo {volume} {82}},\ \bibinfo {pages} {014101}
  (\bibinfo {year} {2010})}\BibitemShut {NoStop}%
\bibitem [{\citenamefont {Sorella}\ \emph {et~al.}(2011)\citenamefont
  {Sorella}, \citenamefont {Casula}, \citenamefont {Spanu},\ and\ \citenamefont
  {Dal~Corso}}]{2011SOR}%
  \BibitemOpen
  \bibfield  {author} {\bibinfo {author} {\bibfnamefont {S.}~\bibnamefont
  {Sorella}}, \bibinfo {author} {\bibfnamefont {M.}~\bibnamefont {Casula}},
  \bibinfo {author} {\bibfnamefont {L.}~\bibnamefont {Spanu}}, \ and\ \bibinfo
  {author} {\bibfnamefont {A.}~\bibnamefont {Dal~Corso}},\ }\href@noop {}
  {\bibfield  {journal} {\bibinfo  {journal} {Phys. Rev. B}\ }\textbf {\bibinfo
  {volume} {83}},\ \bibinfo {pages} {075119} (\bibinfo {year}
  {2011})}\BibitemShut {NoStop}%
\bibitem [{\citenamefont {Becca}\ and\ \citenamefont
  {Sorella}(2017)}]{2017BEC}%
  \BibitemOpen
  \bibfield  {author} {\bibinfo {author} {\bibfnamefont {F.}~\bibnamefont
  {Becca}}\ and\ \bibinfo {author} {\bibfnamefont {S.}~\bibnamefont
  {Sorella}},\ }\href@noop {} {\emph {\bibinfo {title} {{Quantum Monte Carlo
  approaches for correlated systems}}}}\ (\bibinfo  {publisher} {Cambridge
  University Press},\ \bibinfo {year} {2017})\BibitemShut {NoStop}%
\bibitem [{\citenamefont {Giannozzi}\ \emph {et~al.}(2009)\citenamefont
  {Giannozzi}, \citenamefont {Baroni}, \citenamefont {Bonini}, \citenamefont
  {Calandra}, \citenamefont {Car}, \citenamefont {Cavazzoni}, \citenamefont
  {Ceresoli}, \citenamefont {Chiarotti}, \citenamefont {Cococcioni},
  \citenamefont {Dabo} \emph {et~al.}}]{2009GIA}%
  \BibitemOpen
  \bibfield  {author} {\bibinfo {author} {\bibfnamefont {P.}~\bibnamefont
  {Giannozzi}}, \bibinfo {author} {\bibfnamefont {S.}~\bibnamefont {Baroni}},
  \bibinfo {author} {\bibfnamefont {N.}~\bibnamefont {Bonini}}, \bibinfo
  {author} {\bibfnamefont {M.}~\bibnamefont {Calandra}}, \bibinfo {author}
  {\bibfnamefont {R.}~\bibnamefont {Car}}, \bibinfo {author} {\bibfnamefont
  {C.}~\bibnamefont {Cavazzoni}}, \bibinfo {author} {\bibfnamefont
  {D.}~\bibnamefont {Ceresoli}}, \bibinfo {author} {\bibfnamefont {G.~L.}\
  \bibnamefont {Chiarotti}}, \bibinfo {author} {\bibfnamefont {M.}~\bibnamefont
  {Cococcioni}}, \bibinfo {author} {\bibfnamefont {I.}~\bibnamefont {Dabo}},
  \emph {et~al.},\ }\href@noop {} {\bibfield  {journal} {\bibinfo  {journal}
  {J. Phys. Condens. Matter.}\ }\textbf {\bibinfo {volume} {21}},\ \bibinfo
  {pages} {395502} (\bibinfo {year} {2009})}\BibitemShut {NoStop}%
\bibitem [{\citenamefont {Perdew}\ \emph {et~al.}(1996)\citenamefont {Perdew},
  \citenamefont {Burke},\ and\ \citenamefont {Ernzerhof}}]{1996PER}%
  \BibitemOpen
  \bibfield  {author} {\bibinfo {author} {\bibfnamefont {J.~P.}\ \bibnamefont
  {Perdew}}, \bibinfo {author} {\bibfnamefont {K.}~\bibnamefont {Burke}}, \
  and\ \bibinfo {author} {\bibfnamefont {M.}~\bibnamefont {Ernzerhof}},\ }\href
  {\doibase 10.1103/PhysRevLett.77.3865} {\bibfield  {journal} {\bibinfo
  {journal} {Phys. Rev. Lett.}\ }\textbf {\bibinfo {volume} {77}},\ \bibinfo
  {pages} {3865} (\bibinfo {year} {1996})}\BibitemShut {NoStop}%
\bibitem [{\citenamefont {Pavone}\ \emph {et~al.}(1993)\citenamefont {Pavone},
  \citenamefont {Karch}, \citenamefont {Sch\"utt}, \citenamefont {Windl},
  \citenamefont {Strauch}, \citenamefont {Giannozzi},\ and\ \citenamefont
  {Baroni}}]{1993PAV}%
  \BibitemOpen
  \bibfield  {author} {\bibinfo {author} {\bibfnamefont {P.}~\bibnamefont
  {Pavone}}, \bibinfo {author} {\bibfnamefont {K.}~\bibnamefont {Karch}},
  \bibinfo {author} {\bibfnamefont {O.}~\bibnamefont {Sch\"utt}}, \bibinfo
  {author} {\bibfnamefont {W.}~\bibnamefont {Windl}}, \bibinfo {author}
  {\bibfnamefont {D.}~\bibnamefont {Strauch}}, \bibinfo {author} {\bibfnamefont
  {P.}~\bibnamefont {Giannozzi}}, \ and\ \bibinfo {author} {\bibfnamefont
  {S.}~\bibnamefont {Baroni}},\ }\href {\doibase 10.1103/PhysRevB.48.3156}
  {\bibfield  {journal} {\bibinfo  {journal} {Phys. Rev. B}\ }\textbf {\bibinfo
  {volume} {48}},\ \bibinfo {pages} {3156} (\bibinfo {year}
  {1993})}\BibitemShut {NoStop}%
\bibitem [{\citenamefont {Kresse}\ \emph {et~al.}(1995)\citenamefont {Kresse},
  \citenamefont {Furthm{\"u}ller},\ and\ \citenamefont {Hafner}}]{1995KRE}%
  \BibitemOpen
  \bibfield  {author} {\bibinfo {author} {\bibfnamefont {G.}~\bibnamefont
  {Kresse}}, \bibinfo {author} {\bibfnamefont {J.}~\bibnamefont
  {Furthm{\"u}ller}}, \ and\ \bibinfo {author} {\bibfnamefont {J.}~\bibnamefont
  {Hafner}},\ }\href@noop {} {\bibfield  {journal} {\bibinfo  {journal} {EPL}\
  }\textbf {\bibinfo {volume} {32}},\ \bibinfo {pages} {729} (\bibinfo {year}
  {1995})}\BibitemShut {NoStop}%
\bibitem [{\citenamefont {Occelli}\ \emph {et~al.}(2003)\citenamefont
  {Occelli}, \citenamefont {Loubeyre},\ and\ \citenamefont
  {LeToullec}}]{2003OCC}%
  \BibitemOpen
  \bibfield  {author} {\bibinfo {author} {\bibfnamefont {F.}~\bibnamefont
  {Occelli}}, \bibinfo {author} {\bibfnamefont {P.}~\bibnamefont {Loubeyre}}, \
  and\ \bibinfo {author} {\bibfnamefont {R.}~\bibnamefont {LeToullec}},\
  }\href@noop {} {\bibfield  {journal} {\bibinfo  {journal} {Nat. Mater.}\
  }\textbf {\bibinfo {volume} {2}},\ \bibinfo {pages} {151} (\bibinfo {year}
  {2003})}\BibitemShut {NoStop}%
\bibitem [{\citenamefont {McSkimin}\ and\ \citenamefont
  {Andreatch~Jr}(1972)}]{1972MCS}%
  \BibitemOpen
  \bibfield  {author} {\bibinfo {author} {\bibfnamefont {H.}~\bibnamefont
  {McSkimin}}\ and\ \bibinfo {author} {\bibfnamefont {P.}~\bibnamefont
  {Andreatch~Jr}},\ }\href@noop {} {\bibfield  {journal} {\bibinfo  {journal}
  {J. Appl. Phys.}\ }\textbf {\bibinfo {volume} {43}},\ \bibinfo {pages} {2944}
  (\bibinfo {year} {1972})}\BibitemShut {NoStop}%
\bibitem [{\citenamefont {Vinet}\ \emph {et~al.}(1987)\citenamefont {Vinet},
  \citenamefont {Smith}, \citenamefont {Ferrante},\ and\ \citenamefont
  {Rose}}]{1987VIN}%
  \BibitemOpen
  \bibfield  {author} {\bibinfo {author} {\bibfnamefont {P.}~\bibnamefont
  {Vinet}}, \bibinfo {author} {\bibfnamefont {J.~R.}\ \bibnamefont {Smith}},
  \bibinfo {author} {\bibfnamefont {J.}~\bibnamefont {Ferrante}}, \ and\
  \bibinfo {author} {\bibfnamefont {J.~H.}\ \bibnamefont {Rose}},\ }\href
  {\doibase 10.1103/PhysRevB.35.1945} {\bibfield  {journal} {\bibinfo
  {journal} {Phys. Rev. B}\ }\textbf {\bibinfo {volume} {35}},\ \bibinfo
  {pages} {1945} (\bibinfo {year} {1987})}\BibitemShut {NoStop}%
\bibitem [{\citenamefont {Sorella}\ and\ \citenamefont
  {Capriotti}(2010)}]{2010SOR}%
  \BibitemOpen
  \bibfield  {author} {\bibinfo {author} {\bibfnamefont {S.}~\bibnamefont
  {Sorella}}\ and\ \bibinfo {author} {\bibfnamefont {L.}~\bibnamefont
  {Capriotti}},\ }\href@noop {} {\bibfield  {journal} {\bibinfo  {journal} {J.
  Chem. Phys.}\ }\textbf {\bibinfo {volume} {133}},\ \bibinfo {pages} {234111}
  (\bibinfo {year} {2010})}\BibitemShut {NoStop}%
\bibitem [{Note1()}]{Note1}%
  \BibitemOpen
  \bibinfo {note} {2 $\times $ 2 $\times $ 2 conventional supercell containing
  64 atoms in the simulation cell is large enough for obtaining a phonon
  dispersion almost consistent with experimental ones, as shown in Fig.~\ref
  {one-body-dft}~\cite {foot}. This was confirmed by comparing the phonon
  dispersion obtained by the finite displacement method and that obtained by
  the linear-response method with very dense $k$ and $q$ grids at the DFT
  level.}\BibitemShut {Stop}%
\bibitem [{\citenamefont {Harris}(1995)}]{1995HAR}%
  \BibitemOpen
  \bibfield  {author} {\bibinfo {author} {\bibfnamefont {N.~J.}\ \bibnamefont
  {Harris}},\ }\href@noop {} {\bibfield  {journal} {\bibinfo  {journal} {J.
  Phys. Chem.}\ }\textbf {\bibinfo {volume} {99}},\ \bibinfo {pages} {14689}
  (\bibinfo {year} {1995})}\BibitemShut {NoStop}%
\bibitem [{\citenamefont {Mouhat}\ \emph {et~al.}(2017)\citenamefont {Mouhat},
  \citenamefont {Sorella}, \citenamefont {Vuilleumier}, \citenamefont
  {Saitta},\ and\ \citenamefont {Casula}}]{2017MOU}%
  \BibitemOpen
  \bibfield  {author} {\bibinfo {author} {\bibfnamefont {F.}~\bibnamefont
  {Mouhat}}, \bibinfo {author} {\bibfnamefont {S.}~\bibnamefont {Sorella}},
  \bibinfo {author} {\bibfnamefont {R.}~\bibnamefont {Vuilleumier}}, \bibinfo
  {author} {\bibfnamefont {A.~M.}\ \bibnamefont {Saitta}}, \ and\ \bibinfo
  {author} {\bibfnamefont {M.}~\bibnamefont {Casula}},\ }\href@noop {}
  {\bibfield  {journal} {\bibinfo  {journal} {J. Chem. Theory Comput.}\
  }\textbf {\bibinfo {volume} {13}},\ \bibinfo {pages} {2400} (\bibinfo {year}
  {2017})}\BibitemShut {NoStop}%
\bibitem [{\citenamefont {Morresi}\ \emph {et~al.}()\citenamefont {Morresi},
  \citenamefont {Paulatto}, \citenamefont {Vuilleumier},\ and\ \citenamefont
  {Casula}}]{2020MOR}%
  \BibitemOpen
  \bibfield  {author} {\bibinfo {author} {\bibfnamefont {T.}~\bibnamefont
  {Morresi}}, \bibinfo {author} {\bibfnamefont {L.}~\bibnamefont {Paulatto}},
  \bibinfo {author} {\bibfnamefont {R.}~\bibnamefont {Vuilleumier}}, \ and\
  \bibinfo {author} {\bibfnamefont {M.}~\bibnamefont {Casula}},\ }\href@noop {}
  {}\Eprint {http://arxiv.org/abs/2103.04094} {arXiv:2103.04094} \BibitemShut
  {NoStop}%
\bibitem [{Note2()}]{Note2}%
  \BibitemOpen
  \bibinfo {note} {In the Raman mode, two nearest-neighbor carbon atoms are
  displaced in opposite directions by a distance $u$ from their high-symmetry
  positions~\cite {2007MAE}. The distortion in the [1 0 0] direction was
  employed in this study~\cite {1984VAN}. We calculated VMC forces for two
  distorted structures ($u = 0.03$ Bohr and $u = 0.05$ Bohr) and obtained the
  Raman frequency by fitting the forces with a linear function. We also
  calculated VMC and LRDMC energies for the undistorted structure and two
  distorted structures ($u = 0.03$ Bohr and $u = 0.05$ Bohr), then obtained
  Raman frequencies by fitting the energies with a quadratic
  function.}\BibitemShut {Stop}%
\bibitem [{\citenamefont {Moroni}\ \emph {et~al.}(2014)\citenamefont {Moroni},
  \citenamefont {Saccani},\ and\ \citenamefont {Filippi}}]{2014MOR}%
  \BibitemOpen
  \bibfield  {author} {\bibinfo {author} {\bibfnamefont {S.}~\bibnamefont
  {Moroni}}, \bibinfo {author} {\bibfnamefont {S.}~\bibnamefont {Saccani}}, \
  and\ \bibinfo {author} {\bibfnamefont {C.}~\bibnamefont {Filippi}},\
  }\href@noop {} {\bibfield  {journal} {\bibinfo  {journal} {J. Chem. Theory
  Comput.}\ }\textbf {\bibinfo {volume} {10}},\ \bibinfo {pages} {4823}
  (\bibinfo {year} {2014})}\BibitemShut {NoStop}%
\bibitem [{Note3()}]{Note3}%
  \BibitemOpen
  \bibinfo {note} {Since smaller displacements were employed in this study
  (i.e., 0.03 and 0.05 Bohr) than the previous one, the error bars in phonon
  frequencies at $\Gamma $ point are a bit larger.}\BibitemShut {Stop}%
\bibitem [{\citenamefont {Drummond}\ \emph {et~al.}(2004)\citenamefont
  {Drummond}, \citenamefont {Towler},\ and\ \citenamefont {Needs}}]{2004DRU}%
  \BibitemOpen
  \bibfield  {author} {\bibinfo {author} {\bibfnamefont {N.~D.}\ \bibnamefont
  {Drummond}}, \bibinfo {author} {\bibfnamefont {M.~D.}\ \bibnamefont
  {Towler}}, \ and\ \bibinfo {author} {\bibfnamefont {R.~J.}\ \bibnamefont
  {Needs}},\ }\href {\doibase 10.1103/PhysRevB.70.235119} {\bibfield  {journal}
  {\bibinfo  {journal} {Phys. Rev. B}\ }\textbf {\bibinfo {volume} {70}},\
  \bibinfo {pages} {235119} (\bibinfo {year} {2004})}\BibitemShut {NoStop}%
\bibitem [{\citenamefont {Krukau}\ \emph {et~al.}(2006)\citenamefont {Krukau},
  \citenamefont {Vydrov}, \citenamefont {Izmaylov},\ and\ \citenamefont
  {Scuseria}}]{2006KRU}%
  \BibitemOpen
  \bibfield  {author} {\bibinfo {author} {\bibfnamefont {A.~V.}\ \bibnamefont
  {Krukau}}, \bibinfo {author} {\bibfnamefont {O.~A.}\ \bibnamefont {Vydrov}},
  \bibinfo {author} {\bibfnamefont {A.~F.}\ \bibnamefont {Izmaylov}}, \ and\
  \bibinfo {author} {\bibfnamefont {G.~E.}\ \bibnamefont {Scuseria}},\
  }\href@noop {} {\bibfield  {journal} {\bibinfo  {journal} {J. Chem. Phys.}\
  }\textbf {\bibinfo {volume} {125}},\ \bibinfo {pages} {224106} (\bibinfo
  {year} {2006})}\BibitemShut {NoStop}%
\bibitem [{\citenamefont {Sun}\ \emph {et~al.}(2015)\citenamefont {Sun},
  \citenamefont {Ruzsinszky},\ and\ \citenamefont {Perdew}}]{2015SUN}%
  \BibitemOpen
  \bibfield  {author} {\bibinfo {author} {\bibfnamefont {J.}~\bibnamefont
  {Sun}}, \bibinfo {author} {\bibfnamefont {A.}~\bibnamefont {Ruzsinszky}}, \
  and\ \bibinfo {author} {\bibfnamefont {J.~P.}\ \bibnamefont {Perdew}},\
  }\href {\doibase 10.1103/PhysRevLett.115.036402} {\bibfield  {journal}
  {\bibinfo  {journal} {Phys. Rev. Lett.}\ }\textbf {\bibinfo {volume} {115}},\
  \bibinfo {pages} {036402} (\bibinfo {year} {2015})}\BibitemShut {NoStop}%
\bibitem [{\citenamefont {Hao}\ \emph {et~al.}(2012)\citenamefont {Hao},
  \citenamefont {Fang}, \citenamefont {Sun}, \citenamefont {Csonka},
  \citenamefont {Philipsen},\ and\ \citenamefont {Perdew}}]{2012HAO}%
  \BibitemOpen
  \bibfield  {author} {\bibinfo {author} {\bibfnamefont {P.}~\bibnamefont
  {Hao}}, \bibinfo {author} {\bibfnamefont {Y.}~\bibnamefont {Fang}}, \bibinfo
  {author} {\bibfnamefont {J.}~\bibnamefont {Sun}}, \bibinfo {author}
  {\bibfnamefont {G.~I.}\ \bibnamefont {Csonka}}, \bibinfo {author}
  {\bibfnamefont {P.~H.~T.}\ \bibnamefont {Philipsen}}, \ and\ \bibinfo
  {author} {\bibfnamefont {J.~P.}\ \bibnamefont {Perdew}},\ }\href {\doibase
  10.1103/PhysRevB.85.014111} {\bibfield  {journal} {\bibinfo  {journal} {Phys.
  Rev. B}\ }\textbf {\bibinfo {volume} {85}},\ \bibinfo {pages} {014111}
  (\bibinfo {year} {2012})}\BibitemShut {NoStop}%
\bibitem [{\citenamefont {Peintinger}\ \emph {et~al.}(2013)\citenamefont
  {Peintinger}, \citenamefont {Oliveira},\ and\ \citenamefont
  {Bredow}}]{2013PEI}%
  \BibitemOpen
  \bibfield  {author} {\bibinfo {author} {\bibfnamefont {M.~F.}\ \bibnamefont
  {Peintinger}}, \bibinfo {author} {\bibfnamefont {D.~V.}\ \bibnamefont
  {Oliveira}}, \ and\ \bibinfo {author} {\bibfnamefont {T.}~\bibnamefont
  {Bredow}},\ }\href@noop {} {\bibfield  {journal} {\bibinfo  {journal} {J.
  Comput. Chem.}\ }\textbf {\bibinfo {volume} {34}},\ \bibinfo {pages} {451}
  (\bibinfo {year} {2013})}\BibitemShut {NoStop}%
\bibitem [{Note4()}]{Note4}%
  \BibitemOpen
  \bibinfo {note} {$\kappa ({\protect \bf S})$ = the maximum eigenvalue / the
  minimum eigenvalue of the overlap matrix ($S_{\mu ,\nu } = \mathinner
  {\langle {\psi _{\mu }|\psi _{\nu }}\rangle }$)}\BibitemShut {NoStop}%
\bibitem [{\citenamefont {Attaccalite}\ and\ \citenamefont
  {Sorella}(2008)}]{2008ATT}%
  \BibitemOpen
  \bibfield  {author} {\bibinfo {author} {\bibfnamefont {C.}~\bibnamefont
  {Attaccalite}}\ and\ \bibinfo {author} {\bibfnamefont {S.}~\bibnamefont
  {Sorella}},\ }\href@noop {} {\bibfield  {journal} {\bibinfo  {journal} {Phys.
  Rev. Lett.}\ }\textbf {\bibinfo {volume} {100}},\ \bibinfo {pages} {114501}
  (\bibinfo {year} {2008})}\BibitemShut {NoStop}%
\bibitem [{\citenamefont {Subedi}\ and\ \citenamefont {Boeri}(2011)}]{2011SUB}%
  \BibitemOpen
  \bibfield  {author} {\bibinfo {author} {\bibfnamefont {A.}~\bibnamefont
  {Subedi}}\ and\ \bibinfo {author} {\bibfnamefont {L.}~\bibnamefont {Boeri}},\
  }\href@noop {} {\bibfield  {journal} {\bibinfo  {journal} {Phys. Rev. B}\
  }\textbf {\bibinfo {volume} {84}},\ \bibinfo {pages} {020508} (\bibinfo
  {year} {2011})}\BibitemShut {NoStop}%
\bibitem [{\citenamefont {Casula}\ \emph {et~al.}(2012)\citenamefont {Casula},
  \citenamefont {Calandra},\ and\ \citenamefont {Mauri}}]{2012CAS}%
  \BibitemOpen
  \bibfield  {author} {\bibinfo {author} {\bibfnamefont {M.}~\bibnamefont
  {Casula}}, \bibinfo {author} {\bibfnamefont {M.}~\bibnamefont {Calandra}}, \
  and\ \bibinfo {author} {\bibfnamefont {F.}~\bibnamefont {Mauri}},\
  }\href@noop {} {\bibfield  {journal} {\bibinfo  {journal} {Phys. Rev. B}\
  }\textbf {\bibinfo {volume} {86}},\ \bibinfo {pages} {075445} (\bibinfo
  {year} {2012})}\BibitemShut {NoStop}%
\bibitem [{\citenamefont {Marchi}\ \emph {et~al.}(2009)\citenamefont {Marchi},
  \citenamefont {Azadi}, \citenamefont {Casula},\ and\ \citenamefont
  {Sorella}}]{2009MAR}%
  \BibitemOpen
  \bibfield  {author} {\bibinfo {author} {\bibfnamefont {M.}~\bibnamefont
  {Marchi}}, \bibinfo {author} {\bibfnamefont {S.}~\bibnamefont {Azadi}},
  \bibinfo {author} {\bibfnamefont {M.}~\bibnamefont {Casula}}, \ and\ \bibinfo
  {author} {\bibfnamefont {S.}~\bibnamefont {Sorella}},\ }\href@noop {}
  {\bibfield  {journal} {\bibinfo  {journal} {J. Chem. Phys.}\ }\textbf
  {\bibinfo {volume} {131}},\ \bibinfo {pages} {154116} (\bibinfo {year}
  {2009})}\BibitemShut {NoStop}%
\bibitem [{\citenamefont {Monkhorst}\ and\ \citenamefont
  {Pack}(1976)}]{1976MON}%
  \BibitemOpen
  \bibfield  {author} {\bibinfo {author} {\bibfnamefont {H.~J.}\ \bibnamefont
  {Monkhorst}}\ and\ \bibinfo {author} {\bibfnamefont {J.~D.}\ \bibnamefont
  {Pack}},\ }\href {\doibase 10.1103/PhysRevB.13.5188} {\bibfield  {journal}
  {\bibinfo  {journal} {Phys. Rev. B}\ }\textbf {\bibinfo {volume} {13}},\
  \bibinfo {pages} {5188} (\bibinfo {year} {1976})}\BibitemShut {NoStop}%
\bibitem [{\citenamefont {Dal~Corso}(2014)}]{2014DAL}%
  \BibitemOpen
  \bibfield  {author} {\bibinfo {author} {\bibfnamefont {A.}~\bibnamefont
  {Dal~Corso}},\ }\href@noop {} {\bibfield  {journal} {\bibinfo  {journal}
  {Comput. Mater. Sci.}\ }\textbf {\bibinfo {volume} {95}},\ \bibinfo {pages}
  {337} (\bibinfo {year} {2014})}\BibitemShut {NoStop}%
\bibitem [{\citenamefont {Virtanen}\ \emph {et~al.}(2020)\citenamefont
  {Virtanen}, \citenamefont {Gommers}, \citenamefont {Oliphant}, \citenamefont
  {Haberland}, \citenamefont {Reddy}, \citenamefont {Cournapeau}, \citenamefont
  {Burovski}, \citenamefont {Peterson}, \citenamefont {Weckesser},
  \citenamefont {Bright}, \citenamefont {{van der Walt}}, \citenamefont
  {Brett}, \citenamefont {Wilson}, \citenamefont {Millman}, \citenamefont
  {Mayorov}, \citenamefont {Nelson}, \citenamefont {Jones}, \citenamefont
  {Kern}, \citenamefont {Larson}, \citenamefont {Carey}, \citenamefont {Polat},
  \citenamefont {Feng}, \citenamefont {Moore}, \citenamefont {{VanderPlas}},
  \citenamefont {Laxalde}, \citenamefont {Perktold}, \citenamefont {Cimrman},
  \citenamefont {Henriksen}, \citenamefont {Quintero}, \citenamefont {Harris},
  \citenamefont {Archibald}, \citenamefont {Ribeiro}, \citenamefont
  {Pedregosa}, \citenamefont {{van Mulbregt}},\ and\ \citenamefont {{SciPy 1.0
  Contributors}}}]{2020VIR}%
  \BibitemOpen
  \bibfield  {author} {\bibinfo {author} {\bibfnamefont {P.}~\bibnamefont
  {Virtanen}}, \bibinfo {author} {\bibfnamefont {R.}~\bibnamefont {Gommers}},
  \bibinfo {author} {\bibfnamefont {T.~E.}\ \bibnamefont {Oliphant}}, \bibinfo
  {author} {\bibfnamefont {M.}~\bibnamefont {Haberland}}, \bibinfo {author}
  {\bibfnamefont {T.}~\bibnamefont {Reddy}}, \bibinfo {author} {\bibfnamefont
  {D.}~\bibnamefont {Cournapeau}}, \bibinfo {author} {\bibfnamefont
  {E.}~\bibnamefont {Burovski}}, \bibinfo {author} {\bibfnamefont
  {P.}~\bibnamefont {Peterson}}, \bibinfo {author} {\bibfnamefont
  {W.}~\bibnamefont {Weckesser}}, \bibinfo {author} {\bibfnamefont
  {J.}~\bibnamefont {Bright}}, \bibinfo {author} {\bibfnamefont {S.~J.}\
  \bibnamefont {{van der Walt}}}, \bibinfo {author} {\bibfnamefont
  {M.}~\bibnamefont {Brett}}, \bibinfo {author} {\bibfnamefont
  {J.}~\bibnamefont {Wilson}}, \bibinfo {author} {\bibfnamefont {K.~J.}\
  \bibnamefont {Millman}}, \bibinfo {author} {\bibfnamefont {N.}~\bibnamefont
  {Mayorov}}, \bibinfo {author} {\bibfnamefont {A.~R.~J.}\ \bibnamefont
  {Nelson}}, \bibinfo {author} {\bibfnamefont {E.}~\bibnamefont {Jones}},
  \bibinfo {author} {\bibfnamefont {R.}~\bibnamefont {Kern}}, \bibinfo {author}
  {\bibfnamefont {E.}~\bibnamefont {Larson}}, \bibinfo {author} {\bibfnamefont
  {C.~J.}\ \bibnamefont {Carey}}, \bibinfo {author} {\bibfnamefont
  {{\.I}.}~\bibnamefont {Polat}}, \bibinfo {author} {\bibfnamefont
  {Y.}~\bibnamefont {Feng}}, \bibinfo {author} {\bibfnamefont {E.~W.}\
  \bibnamefont {Moore}}, \bibinfo {author} {\bibfnamefont {J.}~\bibnamefont
  {{VanderPlas}}}, \bibinfo {author} {\bibfnamefont {D.}~\bibnamefont
  {Laxalde}}, \bibinfo {author} {\bibfnamefont {J.}~\bibnamefont {Perktold}},
  \bibinfo {author} {\bibfnamefont {R.}~\bibnamefont {Cimrman}}, \bibinfo
  {author} {\bibfnamefont {I.}~\bibnamefont {Henriksen}}, \bibinfo {author}
  {\bibfnamefont {E.~A.}\ \bibnamefont {Quintero}}, \bibinfo {author}
  {\bibfnamefont {C.~R.}\ \bibnamefont {Harris}}, \bibinfo {author}
  {\bibfnamefont {A.~M.}\ \bibnamefont {Archibald}}, \bibinfo {author}
  {\bibfnamefont {A.~H.}\ \bibnamefont {Ribeiro}}, \bibinfo {author}
  {\bibfnamefont {F.}~\bibnamefont {Pedregosa}}, \bibinfo {author}
  {\bibfnamefont {P.}~\bibnamefont {{van Mulbregt}}}, \ and\ \bibinfo {author}
  {\bibnamefont {{SciPy 1.0 Contributors}}},\ }\href {\doibase
  10.1038/s41592-019-0686-2} {\bibfield  {journal} {\bibinfo  {journal} {Nat.
  Methods}\ }\textbf {\bibinfo {volume} {17}},\ \bibinfo {pages} {261}
  (\bibinfo {year} {2020})}\BibitemShut {NoStop}%
\end{thebibliography}%

\end{document}